\documentclass[a4,useAMS,usenatbib,usegraphicx]{mn2e}
\usepackage{subfig}
\usepackage{lscape}
\usepackage{natbib}
\usepackage{amssymb,amsmath}
\usepackage{fixltx2e}
\usepackage{longtable}
\usepackage[T1]{fontenc}
\title[AGES VII : A Filament With Long HI Streams]{The Arecibo Galaxy Environment Survey VII: A Dense Filament With Extremely Long H\textsc{i} Streams}
\author[R. Taylor, R. F. Minchin,  H. Herbst., J. I. Davies,  R. Rodriguez, C. Vazquez]{R. Taylor$^{1}$,$^{5}$\thanks{Email: rhyst@naic.edu}, R. F. Minchin$^1$, H. Herbst$^2$, J. I. Davies$^3$, R. Rodriguez$^4$, C. Vazquez$^4$\\
$^1$Arecibo Observatory, HC03 Box 53995, Arecibo, Puerto Rico 00612\\
$^2$Department of Astronomy, College of Liberal Arts and Sciences, University of Florida, 211 Bryant Space Science Center, Gainesville,\\ FL 32611-2055, USA\\
$^3$School of Physics \& Astronomy, Cardiff University, Queens Buildings, The Parade, Cardiff CF24 3AA, U.K.\\
$^4$University of Puerto Rico, CUH Postal Station, Department of Physics and Electronics, 100 Highway 908, Humacao, PR 00791-4300\\
$^5$Astronomical Institute of the ASCR, Bo\v cn\'i II 1401, 14100, Prague, Czech Republic}

\begin{document}

\date{November 2013}

\pagerange{\pageref{firstpage}--\pageref{lastpage}} \pubyear{2013}

\maketitle

\label{firstpage}

\begin{abstract}
We present completed observations of the NGC 7448 galaxy group and background volume as part of the blind neutral hydrogen Arecibo Galaxy Environment Survey (AGES). Our observations cover a region spanning 5$\times$4 degrees, over a redshift range of approximately -2,000 $<$ $cz$ $<$ 20,000 km~s$^{-1}$. A total of 334 objects are detected, mostly in three overdensities at $cz$ $\sim$7,500, $cz$ $\sim$9,600 and $cz$ $\sim$ 11,400 km~s$^{-1}$. The galaxy density is extremely high (15 per square degree) and many ($\sim$24\%) show signs of extended H\textsc{i} emission, including some features as much as 800 kpc in projected length. We describe the overall characteristics of this environment : kinematics, typical galaxy colours and mass to light ratios, and substructure. To aid in the cataloguing of this data set, we present a new FITS viewer (\textsc{frelled} : Fits Realtime Explorer of Low Latency in Every Dimension). This  incorporates interactive source cataloguing tools which increase our source extraction speed by approximately a factor of 50.
\end{abstract}


\section{Introduction}

The effects of the local galaxy density, or environment, on galaxy evolution are still not well understood. It is known that galaxies in dense environments are deficient in neutral atomic hydrogen (H\textsc{i}) with respect to galaxies of the same morphological type and diameter as those in the field (\citealt{haynes84}), and that they are also redder (see \citealt{bg06} for a review, and also \citealt{luca11}). It is widely believed that the two effects are closely related, with the H\textsc{i} providing a reservoir of fuel for star formation, which when removed results in a quenching of star formation (e.g. \citealt{vol01}). Understanding how the gas is lost, and any consequent effects on galaxy evolution, requires direct observations of the gas content of a large number of galaxies in different environments.

H\textsc{i} is an excellent tracer of the effects of environment - being typically more extended than the stellar component by about a factor of 1.7 (\citealt{swat}) it is less gravitationally bound to the galaxy and therefore easier to remove. In certain circumstances, very long H\textsc{i} streams may be produced extending for tens or even hundreds of kiloparsecs (e.g. \citealt{chung}, \citealt{vhi21}, \citealt{koop}). These offer direct evidence as to how the environment affects the gas content of galaxies, and in particular can indicate the cause of the gas removal.

Cluster environments, especially Virgo, are well-studied, and there are many observational examples of galaxies which are undergoing gas stripping - in many cases clearly due to ram pressure (e.g. \citealt{chung}, \citealt{ooloo}). Other examples are more controversial, with the 250 kpc H\textsc{i} plume extending from NGC 4254 being variously attributed to harassment (\citealt{haynes07}) and ram pressure stripping (\citealt{ooloo}). Moreover, while a few other very dramatic H\textsc{i} features are known in Virgo (e.g. the 500 kpc feature described in \citealt{koop}), these are relatively rare. \cite{ooloo} convincingly argue that the formation of a 110 kpc H\textsc{i} feature (associated with the gas-depleted spiral NGC 4388) in Virgo results from ram pressure stripping -- however, most gas-depleted spirals do not show such extended H\textsc{i} features (e.g. \citealt{me13}, hereafter paper VI, also \citealt{mythesis}) - despite the evidence from \cite{ooloo} and \cite{kent} that such features may be long-lived ($>$ 10$^{8}$ yr). Additionally, although relatively little work has been done to numerically simulate the formation of very long ($>$ 100 kpc) streams via high-speed interactions, \cite{duc} and \cite{bekki} suggest it may be possible. Our understanding of the formation and survival of such features is therefore, at least in some cases, still incomplete.

Nonetheless, there has been significant progress in understanding the various mechanisms by which the gas may be lost, and the resulting effects on the star formation activity of the galaxies. \cite{treu} describe the importance of mergers, harassment, starvation and ram pressure stripping as a function of clustercentric radius, based on observations, while more recently, \cite{cen} simulate gas loss during infall. \cite{vol12} find that star formation efficiency is generally reduced by gas loss, and \cite{wel} describe how the gas loss (and consequent reduction in star formation) may explain the observed morphology-density relation. Overall, then, a model is now emerging where galaxies entering clusters lose gas (primarily via ram pressure stripping); due to the resulting quenching of star formation, this process may also explain the  morphology-density relation.

What is relatively less well-understood is the effect of the field environment on the galaxies prior to entering the cluster (but for a detailed discussion of this ``pre-processing'' in groups, see \citealt{bg06}), where galaxies may not only lose gas but also acquire more via accretion (\citealt{hallenbeck} also claim that this may be possible on the periphery of clusters, see also \citealt{yoon}). \cite{web} claim an H\textsc{i} bridge between M31 and M33 is an example of accretion in the Local Group (see also \citealt{wolf} for more recent observations and discussion). \cite{voidweb} suggest accretion to explain a system of three void galaxies embedded in a common H\textsc{i} envelope, but these galaxies also show signs of interactions -- it is not always straightforward to distinguish observationally between gas streams formed by gas removal from galaxies and by accretion. There is not yet a consensus as to how important gas accretion may be, but for a recent review see \cite{alm}.

These difficulties aside, the direct detection of H\textsc{i} streams can indicate when a galaxy is experiencing gas loss (or accretion) even when the mass of intergalactic gas is low compared to the galaxy itself (e.g. \citealt{vhi21} describe a long stream from a galaxy which would not be regarded as H\textsc{i} deficient). This relies on having sensitive, large area, fully-sampled H\textsc{i} observations. Two such surveys are presently ongoing at Arecibo : the ALFALFA (Arecibo L-band Fast Arecibo L-band Feed Array) survey, and AGES (Arecibo Galaxy Environment Survey). ALFALFA is a very large area survey, covering about 7,000 square degrees of the Arecibo sky, to an \textit{rms} of 2.2 mJy (\citealt{aa}). AGES adopts a different strategy, covering 16 selected areas, with a total area coverage of 200 square degrees (\citealt{auld}, hereafter Paper I). Though much smaller than ALFALFA, it is also much deeper, with a typical \textit{rms} of 0.7 mJy, giving greater sensitivity to diffuse gas.

Each area selected for observation with AGES has a primary target volume in the relatively nearby Universe, typically with $cz$ $<$ 5,000 km~s$^{-1}$. The survey also includes a large ``blind'' volume component in which we have no pre-selected targets. The full redshift range is -2,000 $<$ cz $<$ 20,000 km~s$^{-1}$ (for part of the field considered here, we also have some coverage out to $cz$ = 45,000 km~s$^{-1}$ but we leave analysis of this higher-redshift region to a future paper).

We previously reported (\citealt{d11}, hereafter paper IV) preliminary results focusing on one of our primary targets : the galaxy NGC 7448 and its associated group. This is a spiral galaxy whose extended H\textsc{i} component shows clear signs of interactions with several other nearby galaxies. Paper IV dealt with our target region in detail, however only a very preliminary overview of the background volume was presented. The H\textsc{i} observations at that stage were not complete, and there was also only partial optical coverage with the Sloan Digital Sky Survey (SDSS) at that time. 

Here we present our completed observations and catalogue for the entire region, together with optical photometry from the SDSS. In particular, we report the serendipitous discovery of a large collection of H\textsc{i} streams detected in three distinct velocity regions in the background volume. In this paper we focus on the source catalogues and give an overview of the properties of the detections. Our primary aim is to characterise the environment, and while we point out a few noteworthy examples of certain features, we leave a detailed analysis of the individual detections to future works.

The rest of this paper is organised as follows. In section 2 we describe the observations and data reduction, in section 3 we give details of our source extraction procedures, section 4 presents our catalogues, while the results of the analysis are described in section 5 and discussed in section 6.

\section{Observations and data reduction}
\label{sec:Obs}

Our observation and data reduction procedures have been previously described in detail in paper I. We here give only a brief summary. Observations of this field began in June 2008 and were completed in November 2011, using the 7-beam ALFA (Arecibo L-band Feed Array) instrument on the Arecibo telescope in spectral line mode. The field observed to full depth spans 5 degrees of R.A. by 4 degrees of declination centred on NGC 7448. Due to ALFA's hexagonal beam arrangement, a small area outside this range is also included at a lower sensitivity. The full spatial range of the data considered here is from 22:49:00 to 23:10:50 in R.A, and from +13:51:00 to +18:06:00 in declination. 

Observations are performed in drift scan mode, where the telescope is driven to the start of the scan position and the sky allowed to drift overhead. The ALFA instrument is rotated before the start of each scan to give an equal space between the drift tracks of the 7 beams. At the start of each scan a signal from a high-temperature noise diode is injected for 5 seconds to provide flux calibration.

At Arecibo a point takes 12 seconds to cross the beam. Scans are staggered by just under half the beam width (1.5$^{\prime}$) so that each beam makes a Nyquist-sampled map. With the 7 beams of ALFA each point is scanned 25 times, giving a total integration time of 300 seconds per point (the total integration time for the entire field is 160 hours). Each beam records data from 2 polarisations every second over 4096 channels, which span a velocity range from approximately -2,000 km~s$^{-1}$ to + 20,000 km~s$^{-1}$. The velocity resolution is equivalent to 5 km~s$^{-1}$, or 10 km~s$^{-1}$ after Hanning smoothing to remove Gibbs ringing from the Milky Way and other bright sources such as RFI (see below).

The data are reduced using the AIPS++ packages \textsc{livedata} (using  pksreader v19.40, pksbandpass v19.54, and pkswriter v19.20) and \textsc{gridzilla} (using pksgridzilla v19.50; for a full description see \citealt{barnes}), developed by the Australia Telescope National Facility (ATNF). \textsc{livedata} is used to estimate and remove the bandpass and calibrate the resultant spectra; \textsc{gridzilla} co-adds the data to produce the final 3D data cubes. Our data are gridded into 1$^{\prime}$ pixels (the Arecibo beam has a FWHM of 3.5$^{\prime}$ at this wavelength) with a channel width of 5 km~s$^{-1}$. The bandpass across the scan is estimated by the median of the data values, which is robust to bright point sources but can give incorrect values in the vicinity of bright extended sources. The data are analysed using the \textsc{miriad} task \textit{mbspect}.

From the \textsc{miriad} task \textit{imstat}, the median noise level - \textit{rms} - reached is approximately 0.7 mJy/beam (after Hanning smoothing). As a sensitivity limit, a 5$\sigma$ detection with a tophat profile and 50 km~s$^{-1}$ velocity width, at an \textit{rms} of 0.7 mJy, would have an H\textsc{i} mass of 4.1$\times$10$^{4}$ M$_{\odot}$ $D^{2}$ where $D$ is the distance in Mpc.

Follow-up observations were performed using the single-pixel L-wide receiver. This uses the position-switching method, with the ON and OFF source times each of 5 minutes. The data were quickly reduced and additional observations taken if required. We used the WAPPs (Wideband Arecibo Pulsar Processors) with 9-level sampling and 1 polarization per board, giving 4096 channels. We used two bandwidth settings, 50 MHz (giving a velocity resolution of 1.25 km~s$^{-1}$) and 25 MHz (0.63 km~s$^{-1}$ resolution). The bandwidth was adjusted to be centred on the velocity of the purported source, allowing us to easily stack the observations for greater sensitivity if necessary.

The follow-up observations were performed on all of the 127 point sources we considered uncertain (see section \ref{sec:Analysis}). These observations typically required 5 minutes on-source. Additionally, we identified extended H\textsc{i} features associated with around 24\% of our detections (see section \ref{sec:streams}). To be certain that these were real, we obtained follow-up observations of 10 of these features. These were selected to be among the most extended features detected, ensuring that our pointings would not be affected by sidelobe contamination from their associated galaxies, and also the weakest (thus if these are detected then the stronger features should all be real). To securely confirm the status of the observed targets, for these observations we spent 10-20 minutes on-source. 

\begin{figure*}
\begin{center}  
  \subfloat[]{\includegraphics[height=55mm]{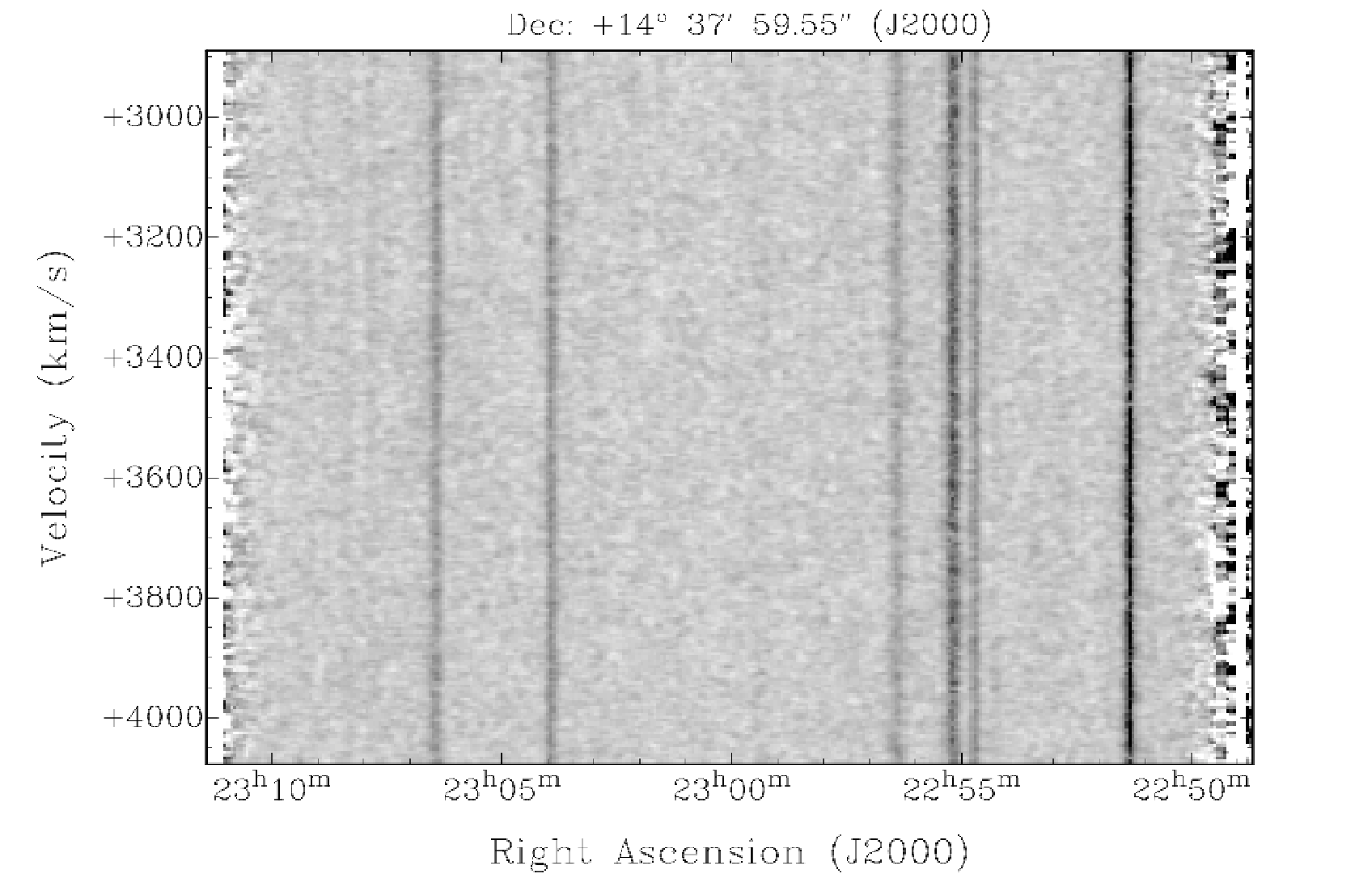}}
  \subfloat[]{\includegraphics[height=55mm]{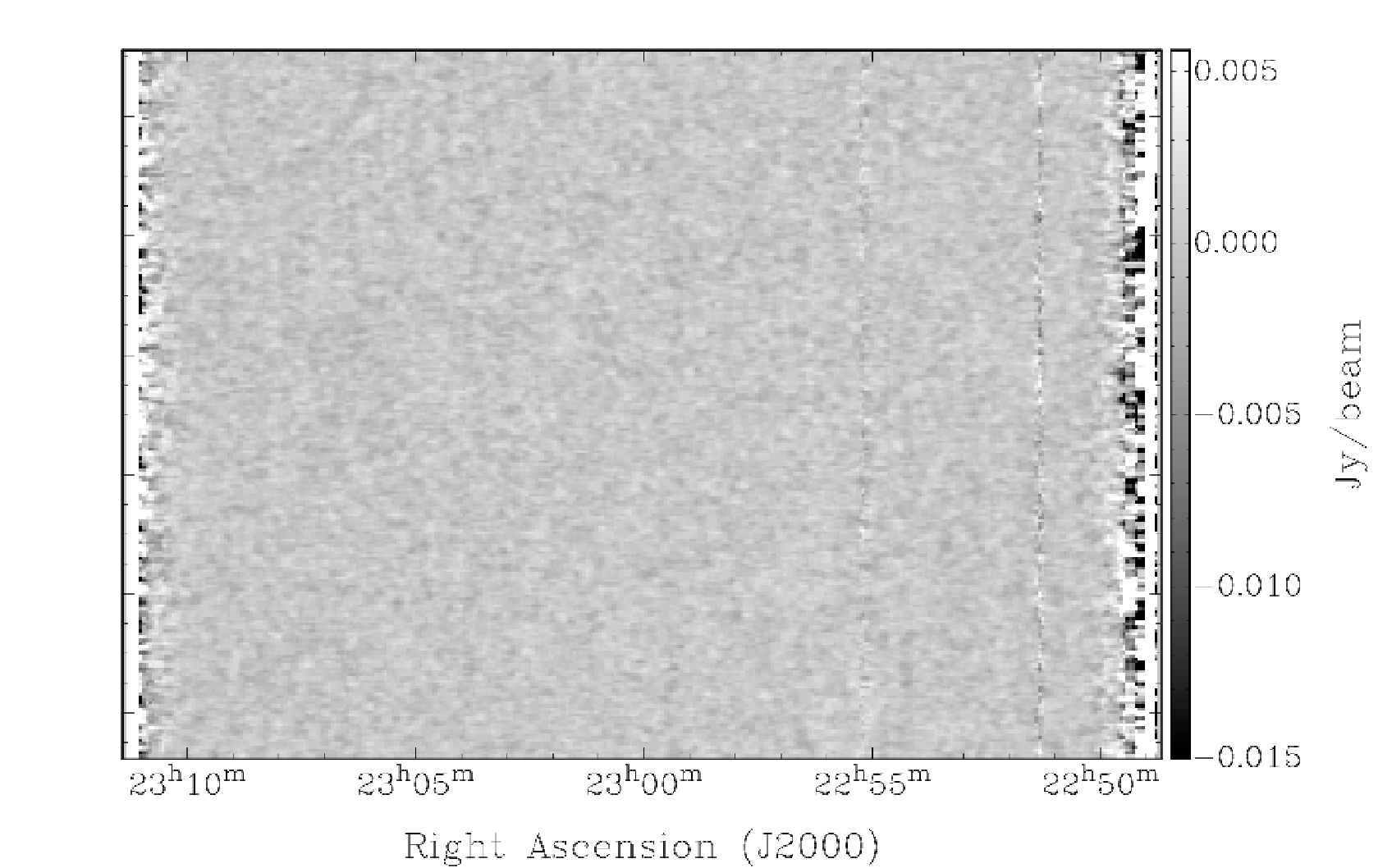}} 
\caption[Baselines]{Position-velocity slices through part of the NGC 7448 data cube before (a) and after (b) a second-order, sigma-clipped polynomial was applied. Each image shows the same region of the cube using the same colour range.}
\label{fig:baselines}
\end{center}
\end{figure*}

Our previous observations of this area, described in paper IV, covered an area of  5.0$\times$2.5 square degrees (of which only 5$\times$2 degrees were to full depth). Our completed observations, described here, cover a significantly larger 5$\times$4 square degrees (entirely to full depth). We also apply an additional baseline correction which significantly improves the quality of the data returned by \textsc{livedata}. Our procedure is to fit and subtract a sigma-clipped second-order polynomial to the data cube, which greatly improves the quality of the data where \textsc{livedata} has problems removing the baseline curvature caused by continuum sources (see figure \ref{fig:baselines}). The larger area, greater sensitivity, better data processing, and improvements to our source extraction (see next section) has resulted in a 90\% increase in the number of H\textsc{i} detections in the whole field (or 38\% in the area which overlaps the area described in paper IV, after accounting for follow-up). 

We note that over the velocity ranges 8,300-8,800 km~s$^{-1}$ and 15,400-15,900 km~s$^{-1}$, sensitivity is greatly reduced owing to RFI from satellites transmitting at the L3 frequency, and the Punta del Este FAA radar, respectively. Our survey is not designed to cope with the extremely bright extended emission from the Milky Way, so we are effectively non-sensitive over the velocity range -100 $<$ $cz$ $<$ +100 km~s$^{-1}$. In this paper we restrict our analysis to extragalactic sources.

\section{Source extraction}
\label{sec:sex}

We use a combination of visual extraction and an automated algorithm, \textsc{glados} (Galaxy Line Analysis for Detection Of Sources). This is fully described in \cite{me12}, hereafter paper V. In brief, it is a signal-to-noise (S/N) based extractor which was specifically designed for AGES data. Each spectrum is searched for S/N values exceeding a user-specified threshold (4$\sigma$), and the corresponding individual polarisations are also searched. Additionally, to minimise the chance of single-channel spikes in the noise from being detected, the candidate detections must have a minimum velocity width (25 km~s$^{-1}$). For full details we refer the reader to paper V.

It is, however, a mistake to describe many source extraction (e.g. \textsc{glados}, \textsc{duchamp}, \textsc{polyfind} - see paper VI for details) as automatic. The major problem is that the reliability of their detections is (except at very high S/N levels) much too low to accept their catalogues at face value. For example the reliability of \textsc{glados} is approximately 16\% at the 4$\sigma$ level, whereas \textsc{duchamp} is closer to 7\% using our data, as we demonstrated in paper VI. The need for a human to first remove any spurious detections from these catalogues means that they are really only semi-automatic.

In comparison, humans are far better able to distinguish between RFI and real sources than some algorithms (see paper VI for a detailed comparison), but suffer from subjectivity and are massively slower than most programs (typically, minutes for an algorithm but days or weeks for a human). We here present a new FITS viewer specifically developed to assist in visual source extraction of H\textsc{i} data cubes (\textsc{frelled} : FITS Realtime Explorer of Low Latency in Every Dimension). We estimate that this has increased visual source extraction speed by a factor $\sim$50. Details are given in appendix \ref{sec:apsource}. The source code is available through our website : www.naic.edu/$\sim$ages. 

We searched the NGC 7448 data cube using both \textit{kvis} and \textsc{frelled} and found that both methods give the same completeness and reliability. One of us (HH) searched the cube using \textit{kvis}, while another (RT) searched with our new viewer. HH found 20 sources that RT did not, and RT found 18 sources that HH did not. Such differences are inevitable given the subjective nature of the procedure; however, RT was able to complete the process in a day, whereas HH required approximately six weeks.
 
We use both visual and automated routines to search the entirety of this data set, excepting the regions affected by RFI where \textsc{glados} is not able to search. Our aim was to detect as many real sources as possible in the whole volume.

\section{Source catalogues}
\label{sec:Analysis}

The source extraction techniques described in section \ref{sec:sex} resulted in an initial catalogue of 425 sources. 127 of these were considered uncertain due to their low S/N level, their detection by only one method or observer, or their lack of optical counterpart (see section \ref{sec:ocs}). We obtained L-wide follow-up observations for all of these. 51 were found to be real and 76 spurious (with observations at the same sensitivity or greater, there is no detection to 3$\sigma$). This leaves us with a final sample of 334 objects (a sample is shown in table \ref{tab:HI} - the full table is available online). Note that in this study we do not include any of the galaxies found within the extensive H\textsc{i} envelope associated with NGC 7448, as we have discussed this complex region in paper IV.

\begin{table*}
\tiny
\begin{center}
\caption[HI data]{Measured parameters of a selected sample of our HI detections. Bracketed values indicate errors following the procedures of \cite{korb}. (1) ID in our catalogue (2,3) RA, Dec in J2000 - errors are the fitting error as described in paper II (4) Heliocentric velocity, km~s$^{-1}$, in  the optical reference frame (5,6) Width at 50 and 20\% of the peak flux, km~s$^{-1}$ (7) Total flux, Jy km~s$^{-1}$ (8) Estimated  MH\textsc{i}/M$_{\odot}$, logarithmic units (9) Peak S/N (10) rms in mJy (11) Optical counterpart status. ``0'' indicates the counterpart has a matching optical redshift, ``1'' that no optical redshift measurement is available. ``A'' indicates other candidates are also present with matching optical redshifts, ``B'' that other candidates do not have optical redshift measurements. ``3'' indicates that the optical counterpart cannot be accurately determined/measured due to the presence of a foreground star or problems with the optical data. (12) Velocity from optical redshift of the most likely optical counterpart, km~s$^{-1}$ (13) Name of most likely optical counterpart in major catalogues. The full table is available online.}
\label{tab:HI}
\begin{tabular}{c c c c c c c c c c c c c}
\hline
\multicolumn{1}{c}{(1)}&       
\multicolumn{1}{c}{(2)}&
\multicolumn{1}{c}{(3)}&
\multicolumn{1}{c}{(4)}&
\multicolumn{1}{c}{(5)}&
\multicolumn{1}{c}{(6)}&
\multicolumn{1}{c}{(7)}&
\multicolumn{1}{c}{(8)}&
\multicolumn{1}{c}{(9)}&
\multicolumn{1}{c}{(10)}&
\multicolumn{1}{c}{(11)}&
\multicolumn{1}{c}{(12)}&
\multicolumn{1}{c}{(13)}\\
ID & R.A. & Dec. & Velocity & W50 & W20 & Flux & MH\textsc{i} & SNPeak & rms & Flag & OptVel & Other Names \\
\hline
  AF7448\_001 & 22:59:35.3(0.7) & 16:46:11(10) & 363(2) & 33(3) & 50(5) & 0.6(0.072) & 6.57 & 45.99 & 0.4 & 1 &  & \\
  AF7448\_005 & 23:06:14.0(0.7) & 14:39:36(10) & 1543(1) & 95(3) & 111(4) & 3.579(0.244) & 8.6 & 77.83 & 0.5 & 0 & 1544 & \\
  AF7448\_016 & 23:05:19.3(0.7) & 16:52:10(10) & 2142(1) & 205(3) & 222(4) & 12.466(0.593) & 9.43 & 138.05 & 0.5 & 0 & 2140 & UGC12350\\
  AF7448\_023 & 22:58:11.9(0.7) & 15:53:15(10) & 4118(2) & 191(3) & 211(5) & 3.496(0.202) & 9.44 & 38.41 & 0.6 & 1B & & \\
  AF7448\_034 & 23:05:23.5(0.7) & 14:59:44(10) & 3938(2) & 177(3) & 195(5) & 3.17(0.187) & 9.36 & 34.52 & 0.6 & 0,3 & 3952 & \\
  AF7448\_056 & 23:01:2.0(0.8) & 14:51:16(12) & 7240(6) & 51(12) & 93(18) & 0.253(0.06) & 8.79 & 7.75 & 0.6 & 0B & 7261 & \\
  AF7448\_073 & 23:05:53.4(0.7) & 16:53:46(10) & 7758(4) & 143(7) & 235(11) & 2.166(0.125) & 9.79 & 33.3 & 0.4 & 0 & 7770 & CGCG453-069\\
  AF7448\_124 & 23:04:14.4(0.7) & 17:53:49(10) & 8704(2) & 92(4) & 112(5) & 1.561(0.131) & 9.74 & 34.3 & 0.6 & 1 &  & UGC12338\\
  AF7448\_138 & 23:00:48.8(0.7) & 16:46:34(10) & 9743(10) & 150(19) & 216(29) & 0.364(0.073) & 9.21 & 5.91 & 0.5 & 1 &  & PGC070251\\
  AF7448\_177 & 23:06:3.4(0.7) & 14:52:33(10) & 10732(3) & 245(6) & 287(9) & 2.22(0.134) & 10.08 & 22.45 & 0.5 & 0A & 10696 & UGC12359\\
\hline
\end{tabular}
\end{center}
\end{table*}

The data are analysed with \textit{mbspect}, as described in \cite{luca}, hereafter paper II. To summarise, the position of the source is found from a Gaussian fit to a moment 0 (integrated flux) map within the galaxy's velocity range. The resultant extracted spectrum is a weighted average of the spectra within a 5$\times$5 pixel box centred on the galaxy. The weighting depends on the beam parameters and distance from the determined source position (assuming a point source). The H\textsc{i} mass is computed using the equation~:
\begin{equation}M_{\rm{HI}} = 2.36\times10^{5}\times d^{2}\times F_{\rm{HI}}\label{HIMass}\end{equation}
Where $M_{\rm{HI}}$ is the H\textsc{i} mass in solar units, $d$ is the distance in Mpc, and $F_{\rm{HI}}$ is the integrated H\textsc{i} flux in Jy km~s$^{-1}$. In table \ref{tab:HI} we assume a point-source to compute the H\textsc{i} flux as we want to measure the H\textsc{i} content of specific galaxies (which have optical diameters that are much less than our 3.5\arcmin beam); for measurements of the extended H\textsc{i} components present around some of our detections, see section \ref{sec:streammass}.

We assume that the galaxies are all in Hubble flow with $H_{0}$ = 71 km~s$^{-1}$. We note, however, that this distance estimate can be subject to considerable uncertainties - we discuss this in detail for AF7448\_001 (at the very low velocity of 363 km~s$^{-1}$) in \cite{me14}. The low-redshift galaxy group associated with NGC 7448, our original target, has twenty redshift-independent distance determinations in NED, but they range from 15.0 - 45.0 Mpc (mean 27.2 Mpc with a standard deviation of 7.1 Mpc). If in pure Hubble flow, NGC 7448 would be at 30.9 Mpc (with $cz$ = 2,194 km~s$^{-1}$), close to the mean redshift-independent distance (the estimated H\textsc{i} mass would differ only by a factor 1.3 using the Hubble-flow distance and mean distance from NED). We therefore here use a Hubble-flow distance estimate for all of our detections, including those at very low redshifts.

The measured parameters (and the derived parameter of mass) are provided in full in our database, available through our website : www.naic.edu/$\sim$ages. We note that in some cases, multiple H\textsc{i} detections are found within the space of a beam, making it impossible to obtain accurate measurements (these objects are flagged in the table with an asterisk).

\subsection{Identification of optical counterparts}
\label{sec:ocs}
This region has been fully observed with the Sloan Digital Sky Survey (SDSS). We visually search the SDSS images (DR9) within a 3.5\arcmin radius of the H\textsc{i} coordinates as determined by $mbspect$. When a candidate is found, we check its measured redshift from the SDSS and NED (note that we check that the NED velocity measurement comes from optical spectroscopy and not previous H\textsc{i} measurements). If the candidate's optical redshift(s) agrees with the H\textsc{i} redshift to within 200 km~s$^{-1}$, the object is defined to have a `sure' optical counterpart (flag value `0' in table \ref{tab:HI}). The choice of 200 km~s$^{-1}$ is somewhat arbitrary, but is designed to be small enough to avoid including any objects which are not really associated with the H\textsc{i} but large enough to account for the typical linewidth of an H\textsc{i} detection. If the optical and H\textsc{i} redshifts differ by more than 200 km~s$^{-1}$ then the optical candidate is rejected. In general the agreement between the H\textsc{i} and optical redshifts is very good, with a mean difference of 20 km~s$^{-1}$.

Unfortunately only 70 of our H\textsc{i} detections have optical counterparts with optical redshift measurements available. For those candidates, we accept them as associated with the H\textsc{i} if they are the only unique candidate within 3.5\arcmin (in most cases their optical position differs from the H\textsc{i} by much less than this, with the mean separation being 22$^{\prime\prime}$, median 16$^{\prime\prime}$). We deem these to be `probable' optical counterparts. Although we treat them as being associated with the H\textsc{i} throughout the analysis, their optical association is given a flag value of `1' in the tables, to indicate the uncertainty.

About 90\% of our detections have optical counterparts within 1$^{\prime}$ of the H\textsc{i} coordinates.  We found only five with H\textsc{i} and optical positions differing by more than 1.5$^{\prime}$. Of these, three are interacting with another source, one is at the edge of the cube and one has a very low S/N. In each case this makes it more difficult for \textit{mbspect} to accurately determine the position of the H\textsc{i} emission. However in all cases, a moment 0 map clearly shows that the H\textsc{i} contours are centred on a position within 1$^{\prime}$ of the purported optical counterpart, so we use these coordinates to identify the optical counterpart rather than those generated by \textit{mbspect}.

\subsection{Optical data analysis}

We perform optical aperture photometry on the SDSS data which is uniformly available in this region, using the tools in the \textsc{ds9} package $funtools$. We restrict photometric measurements to those objects where there is a single sure or probable counterpart, where the H\textsc{i} detection is not in doubt, and where the galaxy's optical emission is not contaminated by any foreground stars (such cases are given a flag value `3' in table \ref{tab:HI}).

The high density of our sample means that in many cases it is simply not possible to determine if the H\textsc{i} really has a unique optical counterpart or is associated with multiple sources. Additionally, some of the optical counterparts are obscured by foreground stars which prevent us from obtaining accurate photometric data. In the analysis which follows we only use those H\textsc{i} detections with a single optical counterpart (unless stated otherwise). This reduces our sample size from 334 H\textsc{i} detections to 209 which we consider to have reliable photometry. A sample of our optical measurements is shown in table \ref{tab:opt}, the full list is available online.

\begin{table*}
\tiny
\begin{center}
\caption[HI data]{Optical parameters for objects with a single optical counterpart. (1) Name in our catalogue, an X indicates the H\textsc{i} may be extended (2,3) RA, Dec in J2000 (4) Morphological type : S = Spiral, I = Irregular, E = Elliptical, M = Merger, P = Peculiar, U = Unknown (5) Separation of optical and H\textsc{i} components, arcminutes (6,7) Apparent $g,i$ magnitudes, corrected for galactic extinction (8,9) Absolute $g,i$ magnitudes, corrected for galactic extinction (10) $g-i$ colour, corrected for galactic extinction (11) H\textsc{i} mass-to-light ratio in the \textit{g} band (12) Physical optical diameter, kpc.}
\label{tab:opt}
\begin{tabular}{c c c c c c c c c c c c}
\hline
  \multicolumn{1}{c}{(1) ID} &
  \multicolumn{1}{c}{(2) RA} &
  \multicolumn{1}{c}{(3) DEC} &
  \multicolumn{1}{c}{(4) M} &
  \multicolumn{1}{c}{(5) Separation} &
  \multicolumn{1}{c}{(6) m$_{g}$} &
  \multicolumn{1}{c}{(7) m$_{i}$} &
  \multicolumn{1}{c}{(8) M$_{g}$} &
  \multicolumn{1}{c}{(9) M$_{i}$} &
  \multicolumn{1}{c}{(10)($g-i$)} &
  \multicolumn{1}{c}{(11)MH\textsc{i}/L$_{g}$} &
  \multicolumn{1}{c}{(12)OptD} \\
\hline		     
 AF7448\_001 & 22:59:35.51 & +16:45:57.4 & I & 0.23 & 17.06 & 16.59 & -11.49 & -11.96 & 0.47 & 0.7 & 0.93\\
 AF7448\_003 & 23:06:59.21 & +17:07:56.9 & I & 0.06 & 18.99 & 18.55 & -13.0 & -13.44 & 0.45 & 1.18 & 3.2\\
 AF7448\_005 & 23:06:15.17 & +14:39:29.0 & I & 0.31 & 16.57 & 16.13 & -15.11 & -15.56 & 0.44 & 2.65 & 4.9\\
 AF7448\_006 & 23:05:11.23 & +14:03:46.6 & I & 0.15 & 16.02 & 15.74 & -15.7 & -15.98 & 0.27 & 0.46 & 4.26\\
 AF7448\_007 & 23:01:17.31 & +18:02:14.2 & U & 0.24 & 17.65 & 16.89 & -14.02 & -14.79 & 0.77 & 2.68 & 3.18\\
 AF7448\_009 & 23:08:33.45 & +18:00:35.0 & I & 0.19 & 18.21 & 17.85 & -14.42 & -14.79 & 0.37 & 3.13 & 8.62\\
 AF7448\_010 & 23:07:44.96 & +17:46:38.1 & I & 0.13 & 17.09 & 16.74 & -15.46 & -15.81 & 0.34 & 1.12 & 7.29\\
 AF7448\_011X & 23:05:58.64 & +17:05:23.9 & S & 0.23 & 16.13 & 15.59 & -16.26 & -16.80 & 0.53 & 1.7 & 15.22\\
 AF7448\_012X & 23:05:08.17 & +17:14:16.1 & S & 0.09 & 16.84 & 16.35 & -15.53 & -16.02 & 0.49 & 0.74 & 7.96\\
 AF7448\_013X & 23:04:34.72 & +17:18:17.4 & I & 0.29 & 17.07 & 16.57 & -15.16 & -15.67 & 0.5 & 1.92 & 9.98\\
\hline
\end{tabular}
\end{center}
\end{table*}

We correct for external extinction using the models of \cite{dust}, via the extinction calculator provided by NED. We note that there is significant variation in the extinction in this region, from 0.1 - 1.0 $g$ magnitudes, and therefore we apply an individual correction to each object.

\section{Results}
\label{sec:Results}
\subsection{The filamentary environment}
\label{sec:fil}
We present a comparison of the detection rates for our previous areas in table \ref{drates} (see papers I-VI). The detection rate in the volume behind the NGC 7448 group is almost double the average of the other background volumes, after subtracting the known clusters.

\begin{table}
\small
\begin{center}
\caption[Detection rates]{Comparison of the detection rates over the background volumes of all AGES data sets to date. Note that NGC 628 (paper I) was a precursor survey with a sensitivity of 1.1 mJy; other fields all reach approximately 0.7 mJy. Columns : (1) Name of field (2) Size of area in square degrees (3) Total number of H\textsc{i} detections (4) No. detections per square degree. In the first part of the table we remove galaxies detected in known clusters, in the second part they are included.}
\label{drates}
\begin{tabular}{c c c c}\\
\hline
  \multicolumn{1}{c}{(1) Field} &
  \multicolumn{1}{c}{(2) Area} &
  \multicolumn{1}{c}{(3) No.} &
  \multicolumn{1}{c}{(4) Rate} \\
\hline
  NGC628 & 5 & 22 & 4.8\\
  NGC1156 & 5 & 38 & 7.6\\
  NGC7332 & 5 & 49 & 9.8\\
  A1367 & 5 & 46 & 9.2\\
  Virgo\_1 & 20 & 193 & 9.7\\
  Virgo\_2 & 5 & 26 & 5.2\\
  Total & 45 & 366 & 8.1\\
\hline
  NGC7448 & 20 & 310 & 15.5\\
  A1367-c & 5 & 97 & 19.4\\
  Virgo\_1-c & 20 & 288 & 14.4\\
  Virgo\_2-c & 5 & 41 & 8.2\\
\hline
\end{tabular}
\end{center}
\end{table}

\begin{figure}
\begin{center}
\includegraphics[width=84mm]{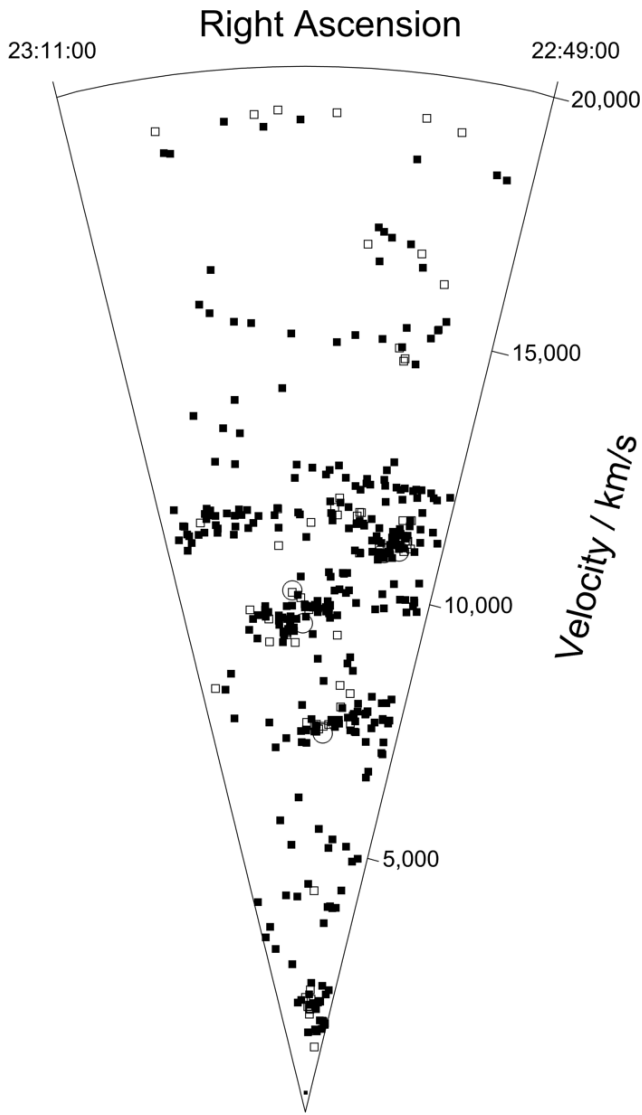}
\caption[Wedge]{Position-redshift diagram for H\textsc{i} detections (filled squares) and non-detections from a NED search (open squares) for the complete volume of the AGES NGC 7448 field. Circles indicate groups as listed in NED.}
\label{fig:wedge}
\end{center}
\end{figure}

\begin{figure*}
\includegraphics[width=180mm]{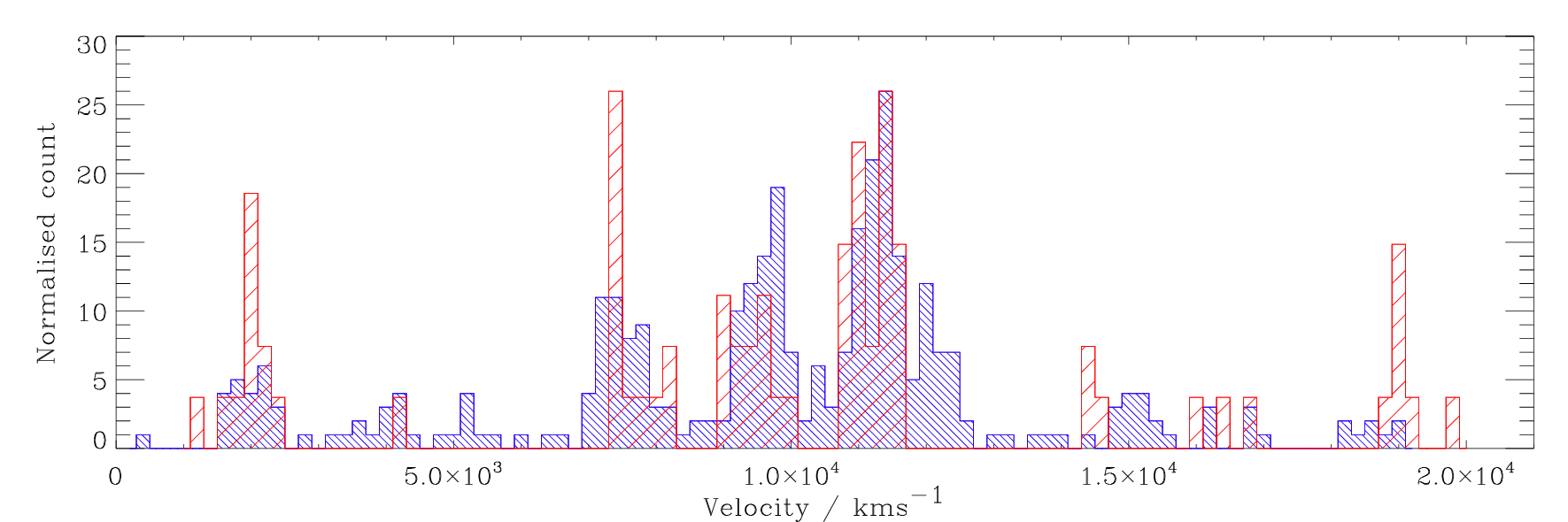}
\caption[Wedge]{Velocity distribution for H\textsc{i} detections (blue, dense hatching) and non-detections from a NED search (red, light hatching) for the complete volume of the AGES NGC 7448 field.}
\label{fig:vdist}
\end{figure*}

A NED search for galaxies in this area (June 2013) reveals 142 objects with measured redshifts below 20,000 km~s$^{-1}$ (in regions where our data are not affected by RFI - see section \ref{sec:Obs}). Of these, only 39 were not detected by AGES (excluding cases where the measured positions and velocities prevent us from distinguishing them from another H\textsc{i} detection). Figure \ref{fig:wedge} compares the large-scale structure in position-velocity space of the H\textsc{i} detections and non-detections, while figure \ref{fig:vdist} compares their velocity distributions. There is no `finger of God' - the classic signature of clusters - but there are three clear overdensities, at around 7,500, 10,000 and 11,000 km~s$^{-1}$. By more than doubling the number of measured redshifts, our survey reveals the structure of this region far more clearly.

In paper II, we noted that clusters at higher redshifts, relative to very nearby clusters like Virgo, may not be detected in H\textsc{i} through wedge diagrams - strong H\textsc{i} deficiency can render galaxies in the cluster core undetectable, so the `finger of God' may not be visible. In this case, however, it does not appear that we have detected a cluster. Not only are the structures traced by both optical (i.e. NED) and the H\textsc{i} very similar, but those structures span at least the full spatial extent of the data cube (5 degrees is equivalent to 12 Mpc at 10,000 km~s$^{-1}$ redshift). No galaxy groups or clusters are detected in this region with the ROSAT X-ray all-sky survey, via a NED search.

\subsection{Extended features in the filamentary environment}
\label{sec:streams}
Motivated by the difficulty in assigning many of the detections unique optical counterparts, we constructed integrated flux maps and renzograms for all of our detections. Integrating the flux over a source increases sensitivity by a factor $\sqrt{n_{c}}$ where $n_{c}$ is the number of channels -- for faint, unresolved sources with high velocity widths this can be particularly important. Renzograms (essentially channel maps) can reveal if there is a positional shift in the peak H\textsc{i} flux as a function of velocity (for instance, if the gas is shared or linked between two optical galaxies at slightly different redshifts). This can help distinguish between objects which are marginally resolved and those with extensions. 

Inspection of these maps revealed that some of the sources may have very extended H\textsc{i} emission (spanning at least two Arecibo beams at the 0.1 Jy beam$^{-1}$ km~s$^{-1}$ level). The flux maps are shown in figures \ref{fig:MomMap1}-\ref{fig:MomMap3}. The figures we show are hybrids of integrated flux maps and renzograms. Here, each image shows the integrated flux contours for many sources. The velocity range over which the flux is integrated depends on the velocity width of each individual source, with colour assigned by its systemic velocity. We leave renzograms of individual sources to a future paper, but note that in most cases the renzograms show at least a hint of the extended emission visible in integrated flux maps.

Integrating the flux over many channels  will not necessarily increase sensitivity by the idealised $\sqrt{n_{c}}$ factor. Baselines are never perfectly flat, and integrating over baseline ripple could give a false positive result. Our follow-up observations re-detected five of the targeted extended sources and failed to detect five others (though two of these suffered from RFI), despite longer integration times (10-20 minutes on-source) and heavy smoothing. These non-detections were among the weakest of the potential streams and likely due to baseline variations.

There are essentially three types of extended H\textsc{i} feature in this sample (see table \ref{tab:streams}). Firstly, there are instances where the H\textsc{i} may not be really extended (or even resolved) at all, but simply appears so due to the superposition of detections at similar velocities within the beam (code `O' in table \ref{tab:streams}). Here it is difficult to accurately disentangle two or more overlapping H\textsc{i} detections and make parameter measurements. While the galaxies may be interacting, we cannot determine exactly how their gas content has been affected (if they are interacting at all). 22 of the galaxies with possible extensions fall into this category, a result of the extremely high (local) density of sources in parts of this volume (for comparison, we found only 6 such cases in the Virgo cubes).

\begin{table}
\tiny
\begin{center}
\caption[Streams]{Basic properties of extended H\textsc{i} features detected in this data set. Columns : (1) ID of galaxies with, or linked by, extended H\textsc{i} emission (see table \ref{tab:HI}, prefix of ID here is AF7448 unless specified otherwise); (2) Total flux in the galaxies assuming point-sources, an asterisk indicates that one or more of the galaxies may be resolved - measurements should be considered highly uncertain in these cases; (3) Total flux present in a volume chosen to contain only those galaxies and any extended emission visible, as described in section \ref{sec:streammass}; (4) Percentage of the total flux present in the extended emission; (5) Description of extended feature, see text.}
\label{tab:streams}
\begin{tabular}{c c c c c}\\
\hline
  \multicolumn{1}{c}{(1) Galaxies} &
  \multicolumn{1}{c}{(2) Flux in} &
  \multicolumn{1}{c}{(3) Total} &
  \multicolumn{1}{c}{(4) Extended} &
  \multicolumn{1}{c}{(5) Code} \\
  & galaxies & flux & fraction & \\
\hline
  011, 012, 013, 016 & 18.3* & 34.0 & 46 & B\\
  014 & 10.9* & 26.7 & 59 & E\\
  035, 036 & 24.2* & 45.0 & 46 & B\\
  046, 054 & 4.4* & 7.3 & 40 & O\\
  050, 052 & 6.9 & 13.7 & 50 & O,S\\
  055, 056 & 1.8 & 2.8 & 36 & O\\
  059, 060 & 7.2 & 9.7 & 26 & B\\
  066, 067 & 2.2 & 3.0 & 27 & O\\
  073 & 2.2 & 5.3 & 58 & E\\
  074 & 0.7 & 1.1 & 36 & E\\
  080 & 1.8 & 3.9 & 54 & E\\
  081, 082 & 2.9 & 5.2 & 42 & O\\
  094, 095 & 1.7 & 2.4 & 31 & O\\
  096, 097, 098, 099 & 10.6 & 14.3 & 26 & B\\
  101, 102, 103, 104 &  &  &  & \\
  106 &  &  &  & \\
  107, 108, 161 & 2.6 & 3.0 & 13 & B\\
  109 & 0.7 & 1.0 & 37 & E\\
  131, 132 & 3.8 & 7.1 & 46 & B\\
  133, 134 & 4.3 & 5.1 & 19 & O\\
  175, 176, 177 & 6.3 & 8.6 & 26 & B\\
  182, 231, 240, 295 & 9.4 & 11.6 & 19 & O,B,E\\
  195, 233 & 2.3 & 3.3 & 30 & O\\
  193, 202, 203, 204 & 6.9 & 9.9 & 30 & O,B\\
  194, 197, 254, 301 & 3.1 & 4.0 & 22 & O,E\\
  208 & 2.2 & 3.2 & 31 & S\\
  210, 211, 297 & 2.2 & 3.7 & 41 & O\\
  216, 217 & 1.3 & 2.2 & 41 & O\\
  218 & 4.4 & 6.6 & 32 & E\\
  221 & 1.3 & 1.6 & 19 & E\\
  223, 224 & 2.7 & 3.1 & 13 & O\\
  230 & 0.9 & 1.4 & 36 & E\\
  245, 252, 253 & 3.0 & 6.5 & 54 & B\\
  256, 257, 258 & 3.5 & 5.6 & 38 & O,B\\
  259, 271, AK019 & 1.9 & 2.2 & 14 & O\\
  264 & 1.2 & 1.9 & 37 & E\\
\hline
\end{tabular}
\end{center}
\end{table}

Secondly there are cases where the H\textsc{i} shows extensions which are significantly larger than the associated detected galaxy, were not detected as separate objects in their own right (usually because they are too faint) but \textit{may} be associated with faint optical sources (code 
`E' in table \ref{tab:streams}). These features have several possible origins - are they simply gas-rich satellites (possibly fuelling their larger companions - see \citealt{sanc} for a review) or does the optical emission result from star formation within streams (e.g. tidal dwarf galaxies or interaction remnants) ?

\begin{figure*}
\begin{center}  
\subfloat[]{\includegraphics[height=84mm]{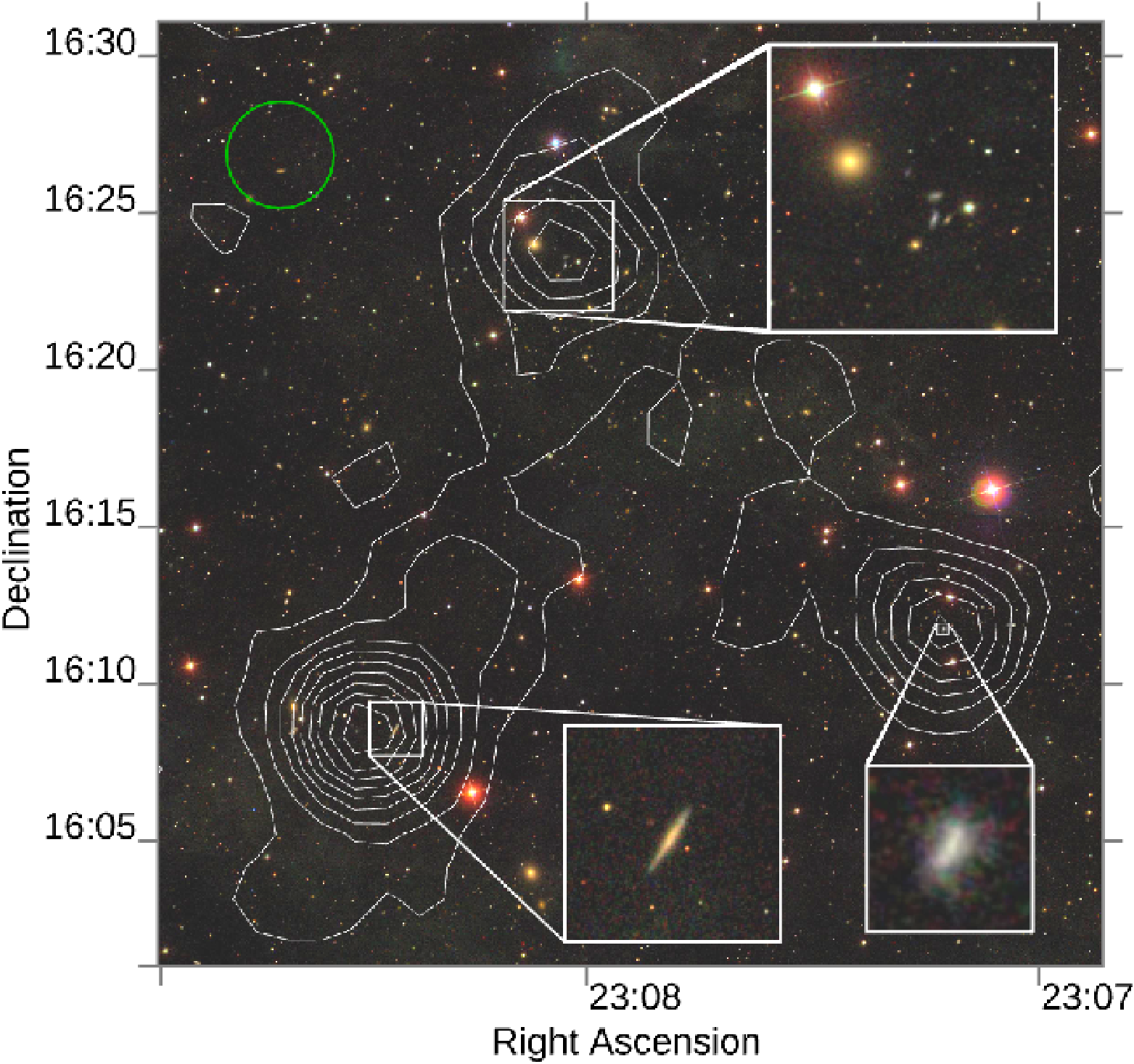}}
\subfloat[]{\includegraphics[height=84mm]{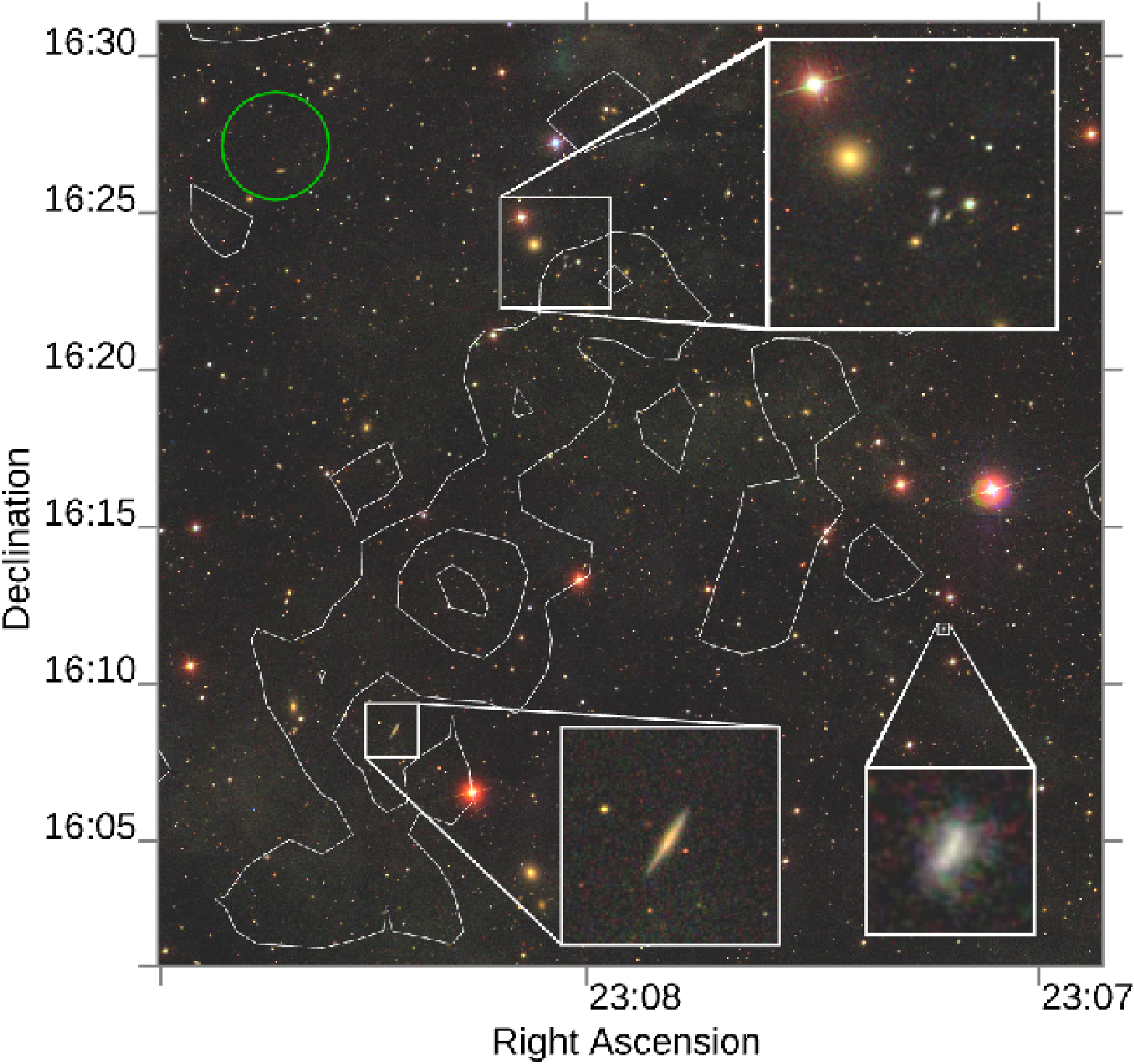}} 
\caption[252Maps]{Integrated flux contours (over the velocity range 11,880 to 12,240 km~s$^{-1}$) of AF7448\_252 (north) which is linked to an edge-on disc (AF7448\_245) approximately 16' south (projected length approximately 800 kpc at a distance of 170 Mpc). A third object to the west (AF7448\_253) may also be interacting with the system. Optical counterparts of individual detections within the stream are shown in the inset images (RGB images from the SDSS). Contours are from 0.1 to 1.0 in steps of 0.1 Jy/beam km~s$^{-1}$. The green circle indicates the 3.5$^\prime$ Arecibo beam. The left panel shows the standard map; the right panels shows the map after removing 3.5$^\prime$ FWHM Gaussians from the three sources, using the H\textsc{i} centroids found via \textit{mbspect}.}
\label{fig:252map}
\end{center}
\end{figure*}

Thirdly, other extended features really appear to be occupy intergalactic space with no optical component - either as streams with only one end terminating at the position of an optically bright galaxy, (code `S' in table \ref{tab:streams}) or bridges between optical galaxies which are well-separated (code `B' in table \ref{tab:streams}). In some cases of individual galaxies or galaxy pairs, the extended emission is huge, as much as 800 kpc in length (see figure \ref{fig:252map}), measuring from the optical centres of the component galaxies. These detections are sufficiently extended that we can accurately determine the H\textsc{i} content of the streams, allowing us to estimate how much damage the interactions between the galaxies has done (see section \ref{sec:streammass}). 

\subsection{Estimating the gas content of the streams}
\label{sec:streammass}
\subsubsection{Beam correction}
In paper IV we described the H\textsc{i} envelope associated with the NGC 7448 group. We estimated the mass of this feature by extracting a subset data cube containing the group and examining its flux value distribution. The negative flux (and low values of positive flux) values followed a Gaussian distribution, which we were able to fit and subtract to estimate the excess present. We measured the total flux (by summing the flux values in each pixel and correcting for the beam shape) and subtracted our measurement of the total flux in the galaxies (i.e. our point-source measurements) to estimate the gas content of the extended H\textsc{i} streams.

AGES data are ordinarily gridded into 1\arcmin square pixels, whereas the beam is in reality circular with a 3.5\arcmin diameter, so the pixels are not independent. This makes it important to correct for the beam shape. Previously we made the assumption that the beam has a tophat profile, with the beam volume therefore given by :
\begin{equation}V_{beam} = \pi\times\left( \frac{FWHM}{2} \right)^{2}\label{circbeam}\end{equation} 
However, a better approximation is to assume a Gaussian shape for the beam (real sources have radial profiles which are indeed Gaussian). In this case we must correct for the volume of the Gaussian :
\begin{equation}V_{beam} = 2\times \pi \times \sigma ^{2} \label{gaussvol}\end{equation}
Where we assume that the beam is circular with $\sigma$ relating to the FWHM by the relation :
\begin{equation}\sigma = \frac{FWHM}{2\sqrt{2\times\ln{2}}}\label{sigfhm}\end{equation}
This correction reduces the measured flux by a factor of $\approx${1.44} compared to using a tophat beam shape. We emphasise that this result only changes our result from paper IV regarding extended emission; our analysis of point sources (using \textit{mbspect}) is not affected.

\subsubsection{Bandpass correction}
To further increase the accuracy of our extended flux estimate, it is also important to correct for the bandpass across the scan. Ordinarily the bandpass is corrected by subtracting the median of the scan. While the median is far more robust to a small number of outliers (such as a bright unresolved galaxy) than the mean, it can be significantly affected by bright extended emission. This results in strong negative `shadows' either side (in the scan direction, in our case Right Ascension - see \citealt{minch}, hereafter paper III, for an example) of bright extended sources, since here the flux along the whole scan will be over-subtracted.

In paper III we applied the ``MinMed'' bandpass estimator (\citealt{put}) to prevent over-subtraction of the bandpass. This splits the scan into 5 segments, finds the median of each and subtracts the minimum of the medians from the entire scan. The ``MedMed'' estimator is very similar, but uses the median of medians instead. We use both of these estimators (separately) but apply them to the reduced cube. Since the \textit{rms} does not vary between regions affected and unaffected by bandpass over-subtraction, this approach should yield essentially the same results as if done as part of the data reduction process.

We find that neither MinMed nor MedMed make a significant difference to the flux estimates of the individual galaxies in the NGC 7448 H\textsc{i} complex ($<$ 5\%), however, the choice of which estimator to use dramatically alters the estimate of the extended H\textsc{i} : applying the Gaussian beam correction described above, we find a total flux of 201 Jy km~s$^{-1}$ using the median, 232 Jy km~s$^{-1}$ using MedMed, and 315 Jy km~s$^{-1}$ using MinMed. MinMed suffers from using a much smaller part of the bandpass and generates visible differences in the bandpasses from channel to channel. We therefore use only the MedMed estimator in our analysis.

\subsubsection{Gas content estimates}
\label{sec:gascont}
In paper IV we estimated that within the NGC 7448 group the gas content of the streams exceeds that of the galaxies by a factor of 2.5. Re-analysing the data with the above improvements reduces this to a factor of 1.6. This is in very good agreement with the earlier result of a factor of 1.5 found by \cite{Wim}, which also further supports our choice to use the MedMed rather than MinMed estimator.

We repeat this analysis for the streams we detect in the background volume. We restrict this to cases where there is a clear H\textsc{i} bridge between two well-separated ($>$ 7$^\prime$ between centres) detections where each shows a single clear optical counterpart. Objects with disturbed H\textsc{i} where we cannot readily identify a plausible cause of the disturbance are ignored - in these cases we cannot determine the gas content of the source galaxies for comparison. Where we identify multiple galaxies linked by H\textsc{i} bridges, we create sub-cubes large enough to contain the entire system. We create a total of 15 subset cubes, containing 53 galaxies. Continuum sources, where present, are masked.

There is a great deal of variation in the gas content external to galaxies, ranging from 10-60\% of the total gas content of each system (mean of 34\% with a standard deviation of 13\%). In the NGC 7448 group about 60\% of the total gas is external to the galaxies, which is high relative to most of the other systems examined but not exceptionally so.

\subsubsection{H\textsc{i} deficiency}
\label{sec:histreamdef}
H\textsc{i} deficiency is a measure of how much gas a galaxy has lost compared to a field galaxy of similar morphology and optical diameter, and is defined as :
\begin{equation}D_{\rm{HI}}= \log(M_{{\rm{HI}}_{ref}}) - \log(M_{{\rm{HI}}_{obs}})\label{DefEqt}\end{equation}
Where $D_{\rm{HI}}$ is the HI deficiency, $M_{{HI}_{ref}}$ is the HI mass of the reference galaxy, and  $M_{{HI}_{obs}}$ is the H\textsc{i} mass of the observed galaxy. A galaxy which has lost 60\% of its gas would have a deficiency of 0.4 -- which, owing to the intrinsic scatter of the H\textsc{i} content of isolated galaxies, would at most be described as borderline deficient. Therefore deficiency measurements cannot help establish if the gas in the observed streams originated within the galaxies, or shows ongoing accretion. For this reason we do not include deficiency measurements in our analysis.

\subsection{Kinematics}

\begin{figure}
\begin{center}
\includegraphics[width=84mm]{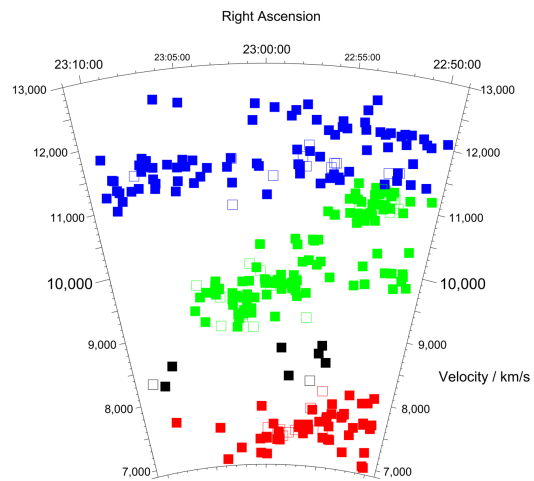}
\caption[BigMap]{Wedge diagram of the highest density region of the cube. Filled squares are H\textsc{i} detections, open squares are NED objects not detected in H\textsc{i}. Colour indicates our (subjectively) assigned population membership : red for population A, green for B, blue for C and black is uncertain.}
\label{fig:zoomwedge}
\end{center}
\end{figure}

We identify three overdensities (by eye) ranging in velocity from 7,000-8,200 km~s$^{-1}$, 9,000-11,200 km~s$^{-1}$ and 11,200-12,600 km~s$^{-1}$ (a close-up wedge diagram is shown in figure \ref{fig:zoomwedge}). We will refer to these associations as populations A, B and C respectively. 

To give a physically meaningful estimate of the velocity dispersion within each population, it is important to account for the large-scale velocity gradient and sub-structures which are present. We use the IDL task \textit{sfit} to fit and subtract a velocity surface from the galaxies, analysing each population separately. For population A galaxies this gives velocity dispersion of 172 km~s$^{-1}$. The presence of substructure in the B and C populations is a significant complication. Population B galaxies show some evidence of a bifurcation in the west, with one part extending into population C and another terminating at 10,000 km~s$^{-1}$ (perhaps extending further beyond the spatial edge of the cube). Within population C, there is a dearth of galaxies from 11,700-11,900 km~s$^{-1}$, and the population from 11,900-13,000 km~s$^{-1}$ is much denser in the western half of the cube compared to the eastern half (31 detections in the west compared with 4 in the east).
 
To try to account for the substructure in population B, we consider two sub-samples. One includes the galaxies at higher velocities (which merge into population C) and excludes those at lower velocities in the western part of the cube (the choice of which to include and exclude is, admittedly, very subjective), while the other includes only the lower-velocity galaxies. The first sample gives a velocity dispersion of 238 km~s$^{-1}$ while the second gives 261 km~s$^{-1}$.

Similarly, fitting a surface to the whole of population C gives a velocity dispersion of 435 km~s$^{-1}$. If we consider population C to contain two separate populations (one from 11,100 - 11,700 km~s$^{-1}$, the other from 11,800 - 12,700 km~s$^{-1}$) and subtract a separate velocity surface from each, then we find much lower velocity dispersions of 144 and 138 km~s$^{-1}$.

\begin{figure}
\begin{center}
\includegraphics[width=84mm]{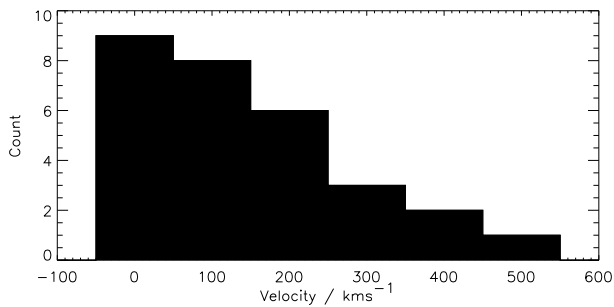}
\caption[RVs]{Distribution of the relative velocities of objects identified as interacting with one another.}
\label{fig:relvels}
\end{center}
\end{figure}

These velocity dispersion estimates indicate that the H\textsc{i} streams we detect could arise from low velocity interactions, despite the high velocity range each population spans. We attempt to estimate the distribution of relative velocities between interacting galaxies. We restrict this analysis to cases where there is a clear H\textsc{i} bridge between two well-separated ($>$ 7$^\prime$ between centres) detections where each shows a single clear optical counterpart. Objects with disturbed H\textsc{i} where we cannot readily identify a plausible cause of the disturbance are ignored. The resulting distribution of relative velocities is shown in figure \ref{fig:relvels}.

The mean relative velocities of the interactors in the sample is 170 km~s$^{-1}$, with a standard deviation of the sample of 185 km~s$^{-1}$. There is a clear bias toward low relative velocities. Potentially this could indicate a selection bias due to the techniques used to reveal the streams, but we do not believe this is the case. Both integrated flux maps and renzograms should reveal if the H\textsc{i} is disturbed in each of our detections, regardless of whether there is an extended feature extending over a significant velocity range. An important caveat is that since we have deliberately avoided cases where we cannot identify both interactors, we may be avoiding high velocity interaction systems where one of the objects is no longer near the disturbed system.

\subsection{Scaling relations of the sample}
\begin{figure*}
\begin{center}  
  \subfloat[]{\includegraphics[width=55mm]{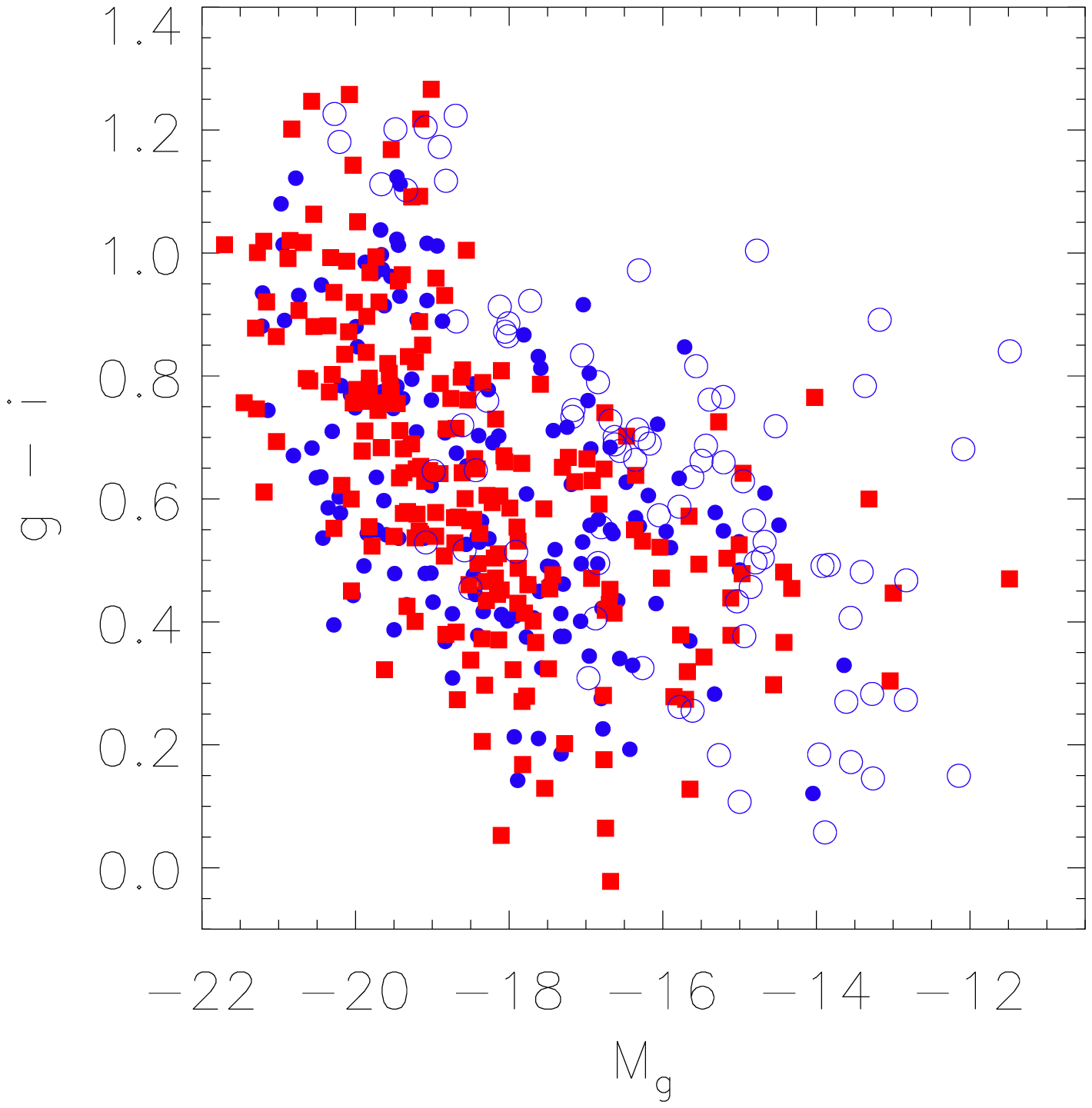}}
  \hspace*{5mm}
  \subfloat[]{\includegraphics[width=55mm]{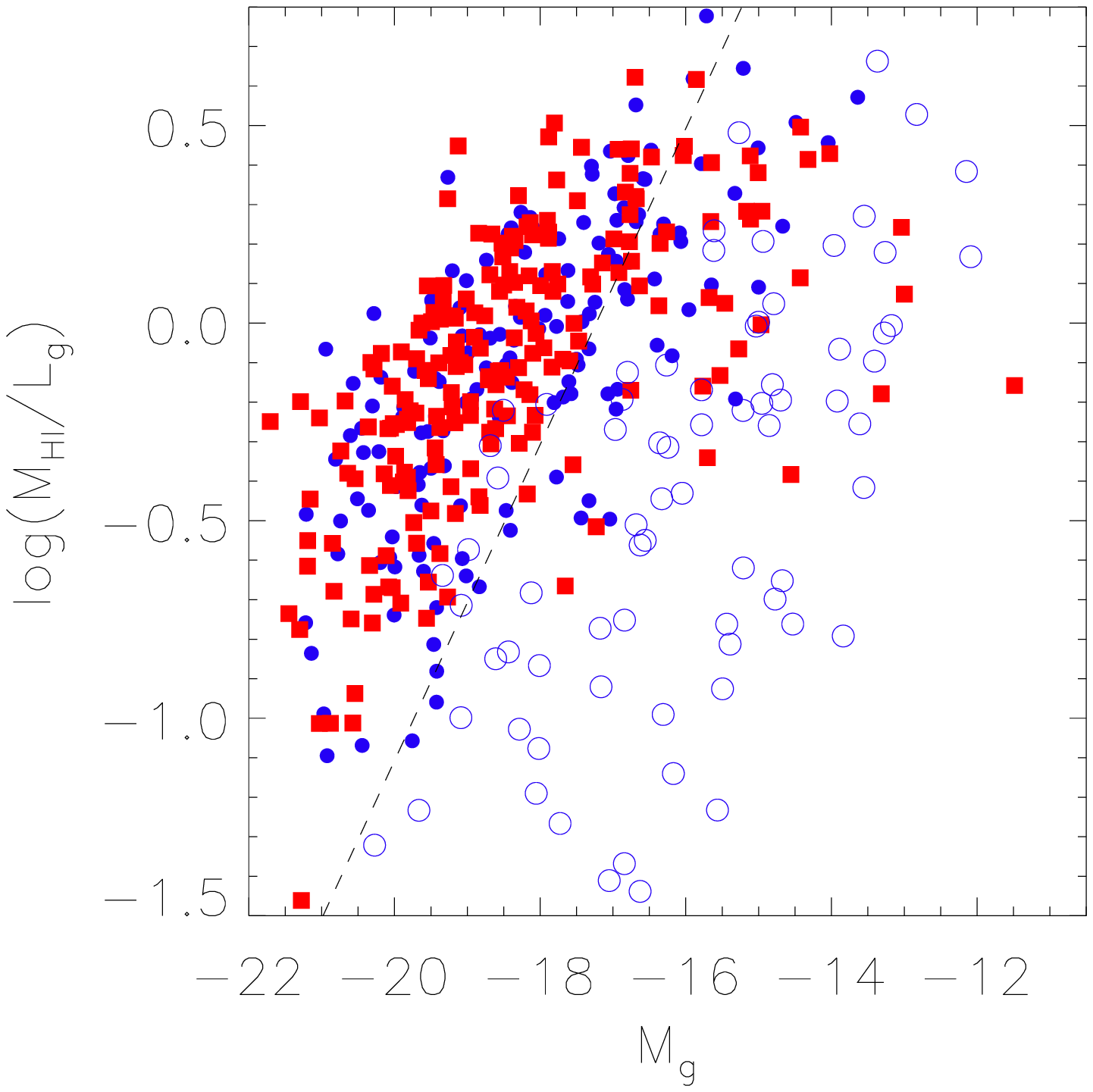}}
  \hspace*{5mm}
  \subfloat[]{\includegraphics[width=55mm]{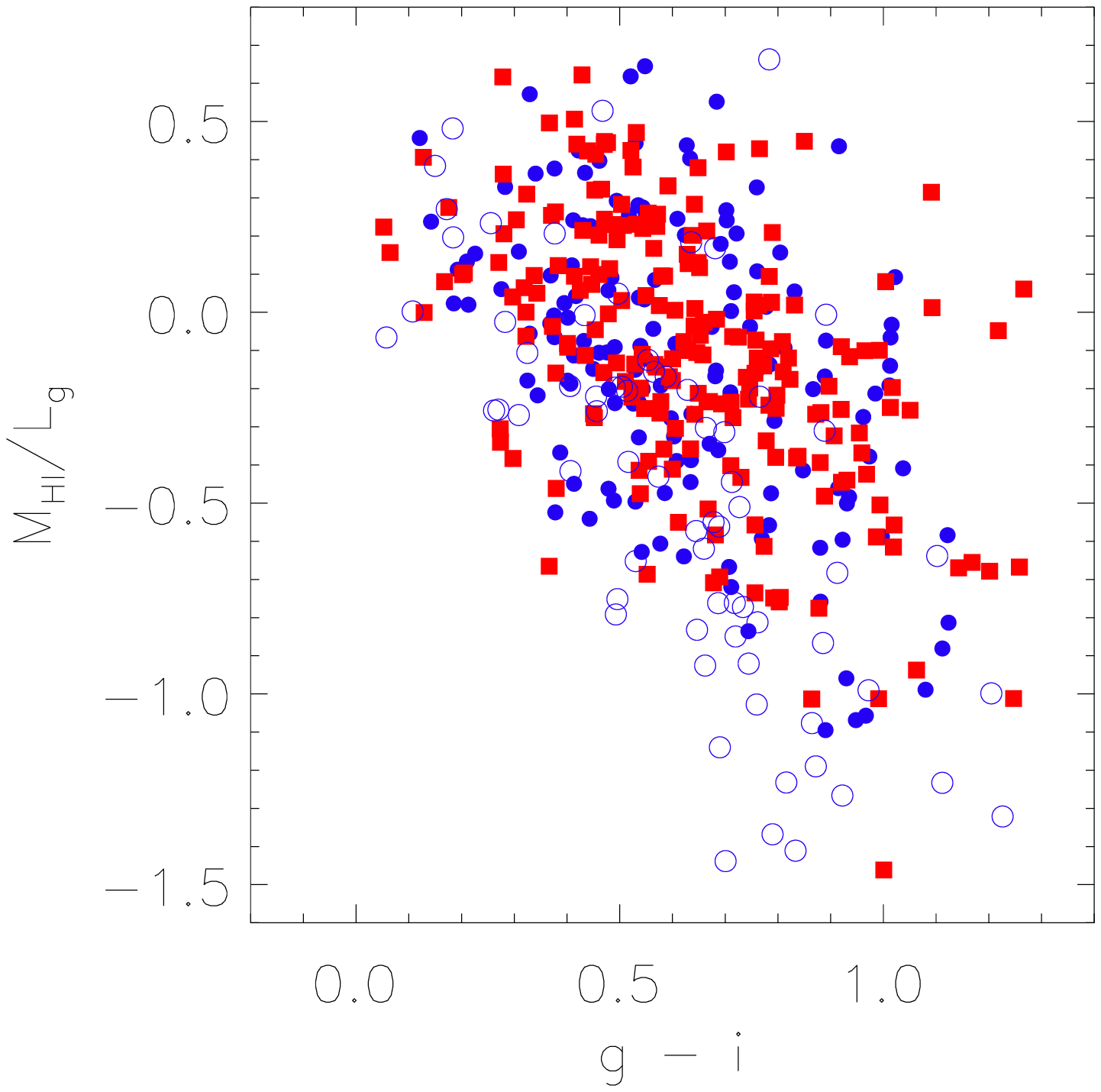}}
\caption[MorphColours]{Major scaling relations for the NGC 7448 sample (filled red squares), the Virgo background sample (filled blue circles) and Virgo cluster sample (open blue circles). (a) Colour-magnitude diagram (b) M$_{HI}$/L$_{g}$ ratio as a function of M$_{g}$. The dashed line shows the sensitivity limit assuming a distance of 160 Mpc for a 5$\sigma$, 50 km~s$^{-1}$ width detection. (c) M$_{HI}$/L$_{g}$ ratio as a function of $g$-$i$ colour. }
\label{fig:scales}
\end{center}
\end{figure*}

We show the major optical properties of our sample in figure \ref{fig:scales}. In figure \ref{fig:scales} (a) we compare the colour-magnitude diagram for this sample with those galaxies detected in and behind the Virgo cluster (see paper V). Galaxies behind NGC 7448 do not appear to be significantly redder than those detected behind the Virgo cluster. A two-dimensional K-S test finds that the populations are different but only at the 2.0$\sigma$ level. In contrast the galaxies within the Virgo cluster itself are clearly redder (at a 7$\sigma$ level from the two-dimensional K-S test).

Figure \ref{fig:scales} (b) shows the variation in M$_{HI}$/L$_{g}$ ratio as a function of absolute magnitude. Again the galaxies in this sample follow the same trend as the galaxies in the less dense environment behind Virgo, whereas the Virgo cluster galaxies clearly show a markedly lower M$_{HI}$/L$_{g}$ ratio. We show the sensitivity limit at 160 Mpc (corresponding to the overdensity at $\sim$11,400 km~s$^{-1}$).

Figure \ref{fig:scales} (c) compares the variation in M$_{HI}$/L$_{g}$ ratio as a function of $g$-$i$ colour. Once again we find a good agreement between this and the Virgo background sample (the K-S test finds the populations are different at only a 2.2$\sigma$ level). Despite the density difference, and the abundance of extended H\textsc{i} features in the 7448 background sample, the two populations are very similar in both their gas contents, colour, and variations of those parameters.

\subsection{Morphologies}
\begin{figure*}
\begin{center}  
  \subfloat[]{\includegraphics[width=55mm]{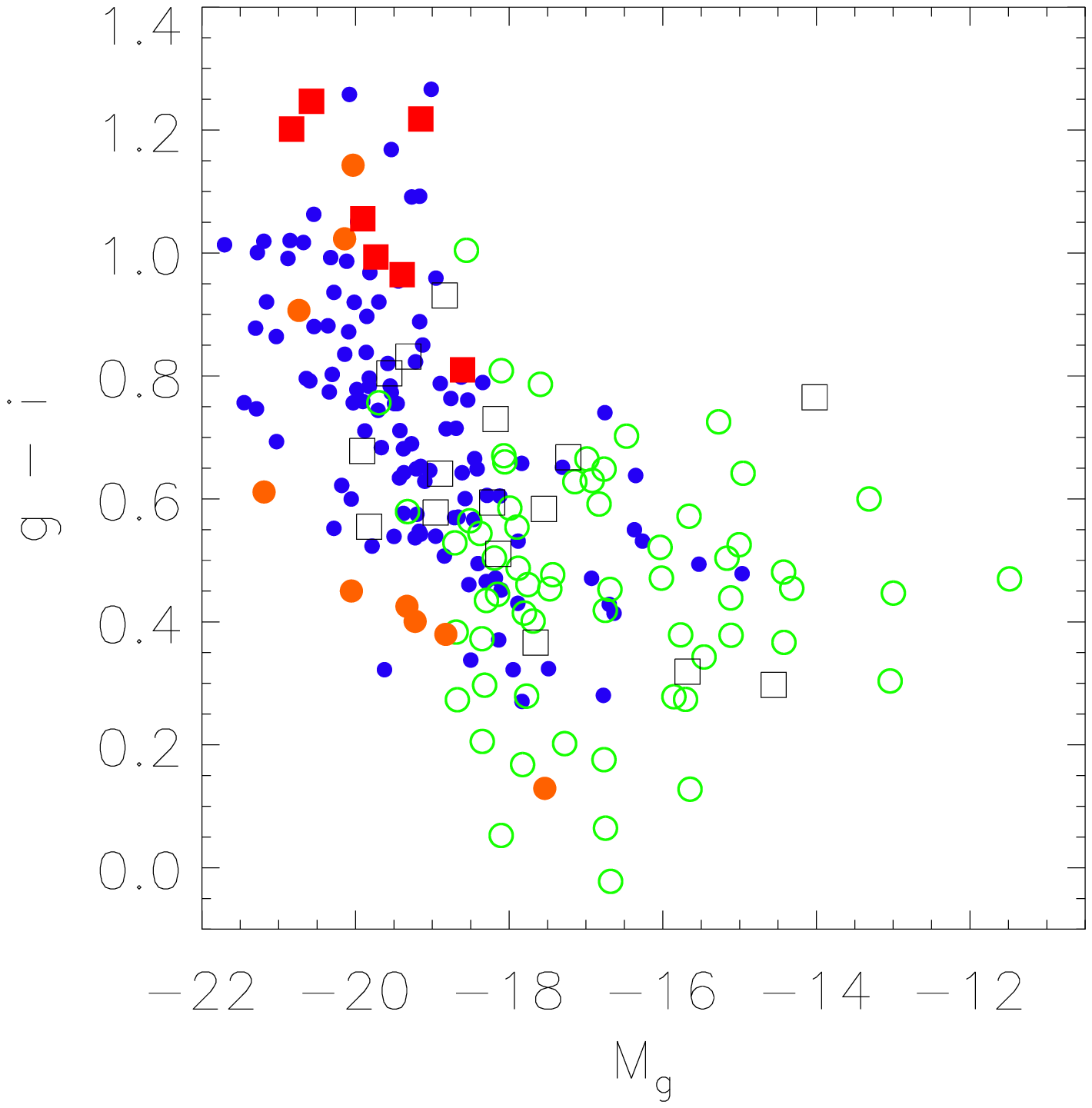}}
  \hspace*{5mm}
  \subfloat[]{\includegraphics[width=55mm]{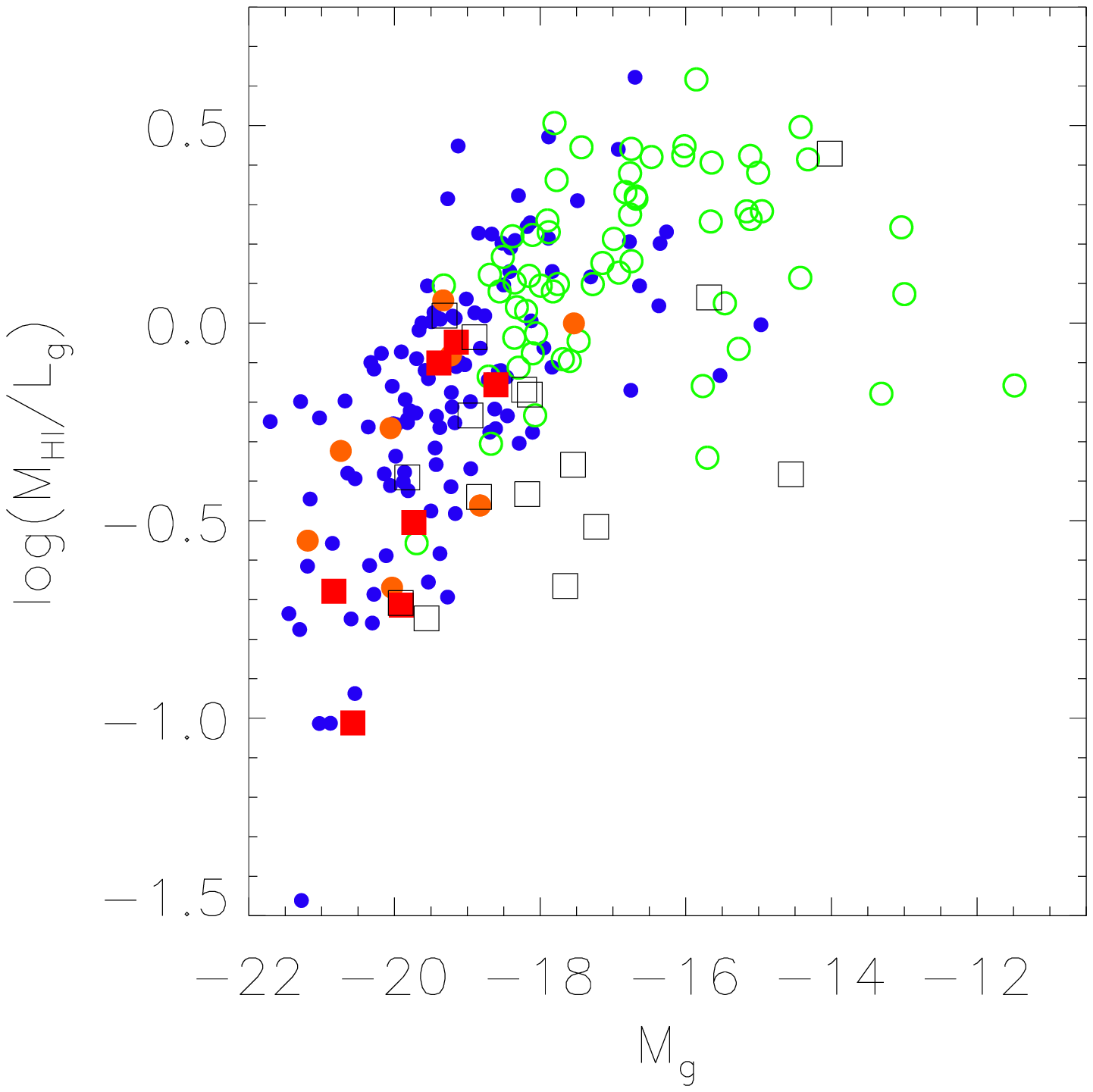}}
  \hspace*{5mm}
  \subfloat[]{\includegraphics[width=55mm]{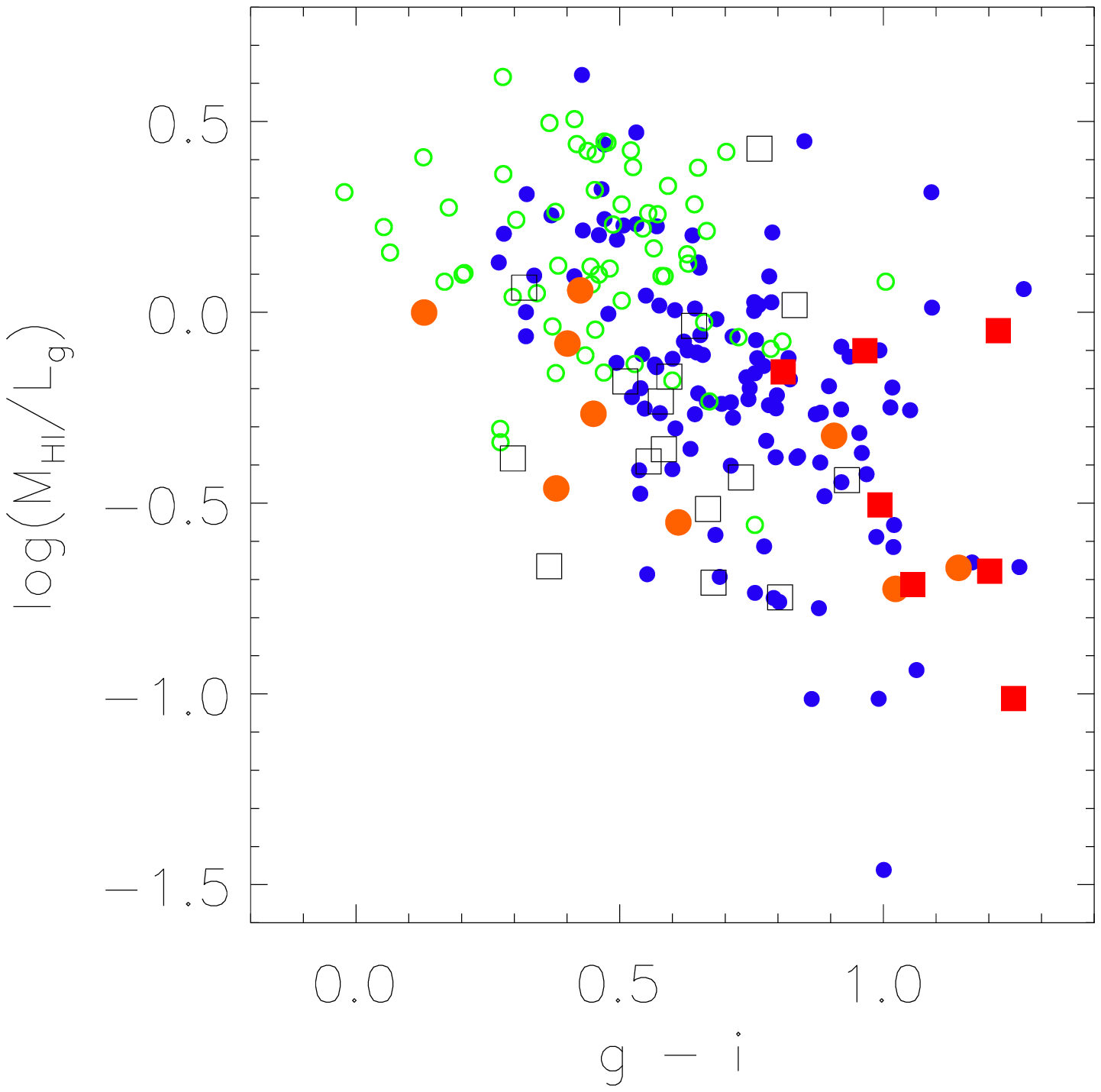}}
\caption[MorphColours]{(a) Colour-magnitude diagram (b) M$_{HI}$/L$_{g}$ ratio as a function of M$_{g}$ (c) M$_{HI}$/L$_{g}$ ratio as a function of $g$-$i$ colour. In all plots, small filled blue circles are spirals, small open green circles are irregulars, large orange filled circles are peculiars/mergers, large filled red squares are early-types, and open black squares are unclassified. Here only galaxies detected in the NGC 7448 data cube are shown.}
\label{fig:morphs}
\end{center}
\end{figure*}

Since there are few existing morphological classifications of the galaxies in this sample, we attempt a rudimentary classification ourselves. We divide the sample into spirals, irregular, early-types, mergers and peculiars, and those without an obvious classification. This is done purely on the basis of a visual inspection of morphology. Though crude, this approach does offer some insights. We show the colour-magnitude, magnitude-gas content, and colour-gas content diagrams with morphological differentiation in figure \ref{fig:morphs}.

The irregulars and spirals define two different but overlapping populations. As we should expect, the spirals are brighter and somewhat less gas-rich than the smaller irregulars. The early-type galaxies are (as expected) generally redder and brighter than the late-types, but they are not always gas poor.

Of the nine objects we classify as mergers or peculiars, six are very blue relative to other detections of similar luminosity, while the other three are relatively red. There is no obvious correlation between their luminosity and M$_{HI}$/L$_{g}$ ratio, however, the six which are relatively blue given their luminosity are also blue compared to galaxies of similar M$_{HI}$/L$_{g}$ ratio.

The low number of peculiar and merging galaxies may seem surprising given the abundance of interactions and density of the environment. It is worth re-emphasising that we have deliberately excluded objects with multiple possible counterparts. One system in particular exemplifies the complexity of the problems when dealing with such cases. AF7448\_202 is centred on a group of three objects, one of which appears to be a previously undiscovered ring galaxy. The H\textsc{i} may be extended to another optically disturbed galaxy (AF7448\_204) about 8$^\prime$ (370 kpc at 160 Mpc distance) away, however, this is difficult to assess. The galaxies in figure \ref{fig:252map} are easy to remove to form the residual map as they are relatively faint and well-separated; here, however, the AF7448\_202 complex is bright and extended. We use a 3.5$^\prime$ FWHM Gaussian for AF744\_204, centred on its optical coordinates, and fit a non-circular Gaussian (4.3 x 4.8$^\prime$, position angle 69$^{\circ}$ measured north through east) using the \textsc{miriad} task \textit{imfit} for  AF7448\_202\footnote{A caveat is that by fitting the three galaxies in this way, we cannot be certain we have not included part of any possible bridge between them and AF744\_204.}. Another nearby source, AF7448\_193 (16$^\prime$ or 760 kpc in projected distance), is an early-type galaxy which shows a possible north-south extension, but it is unclear if this relates to the AF7448\_202 complex.

\begin{figure*}
\begin{center}  
\subfloat[]{\includegraphics[height=84mm]{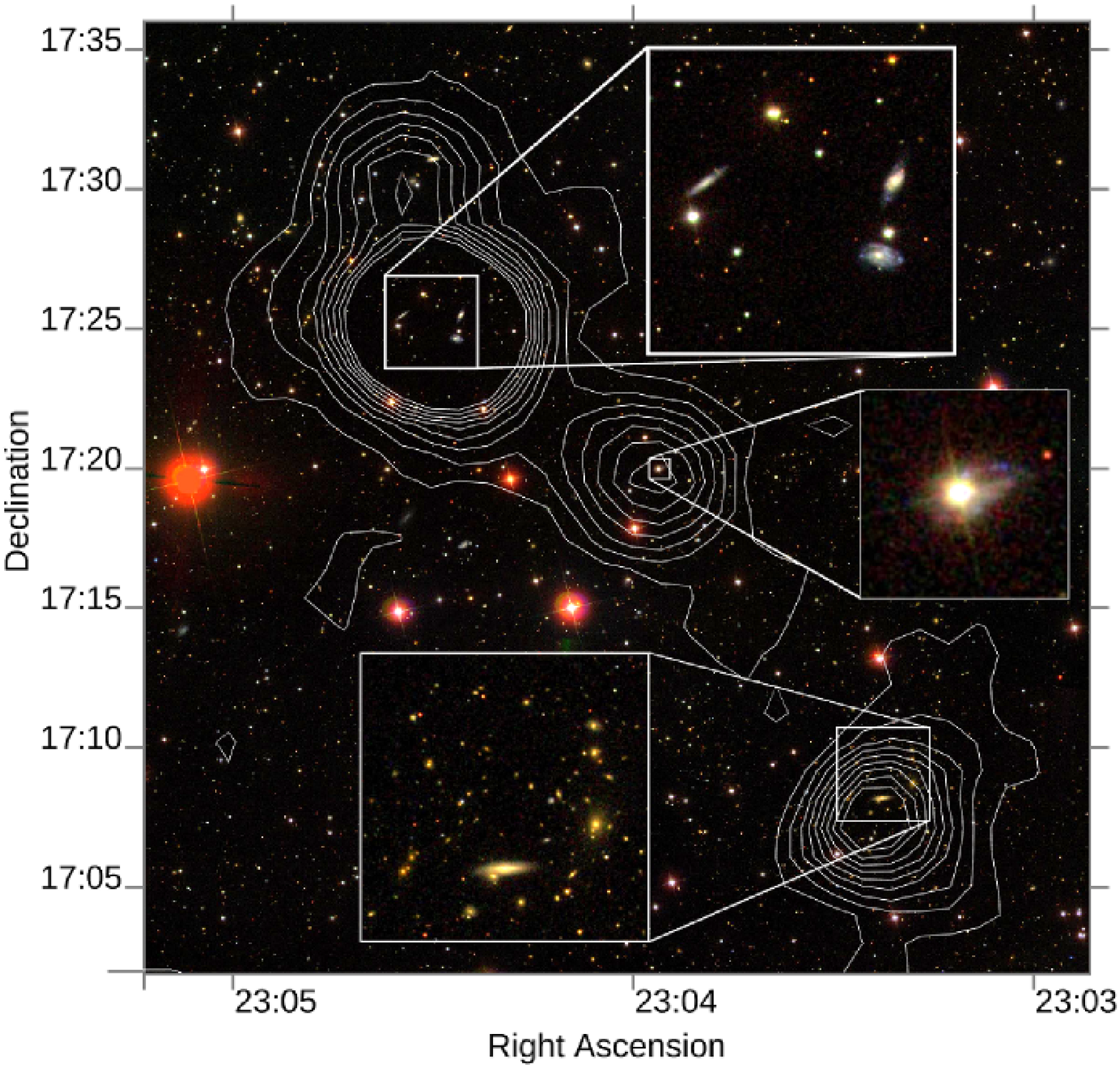}}
\subfloat[]{\includegraphics[height=84mm]{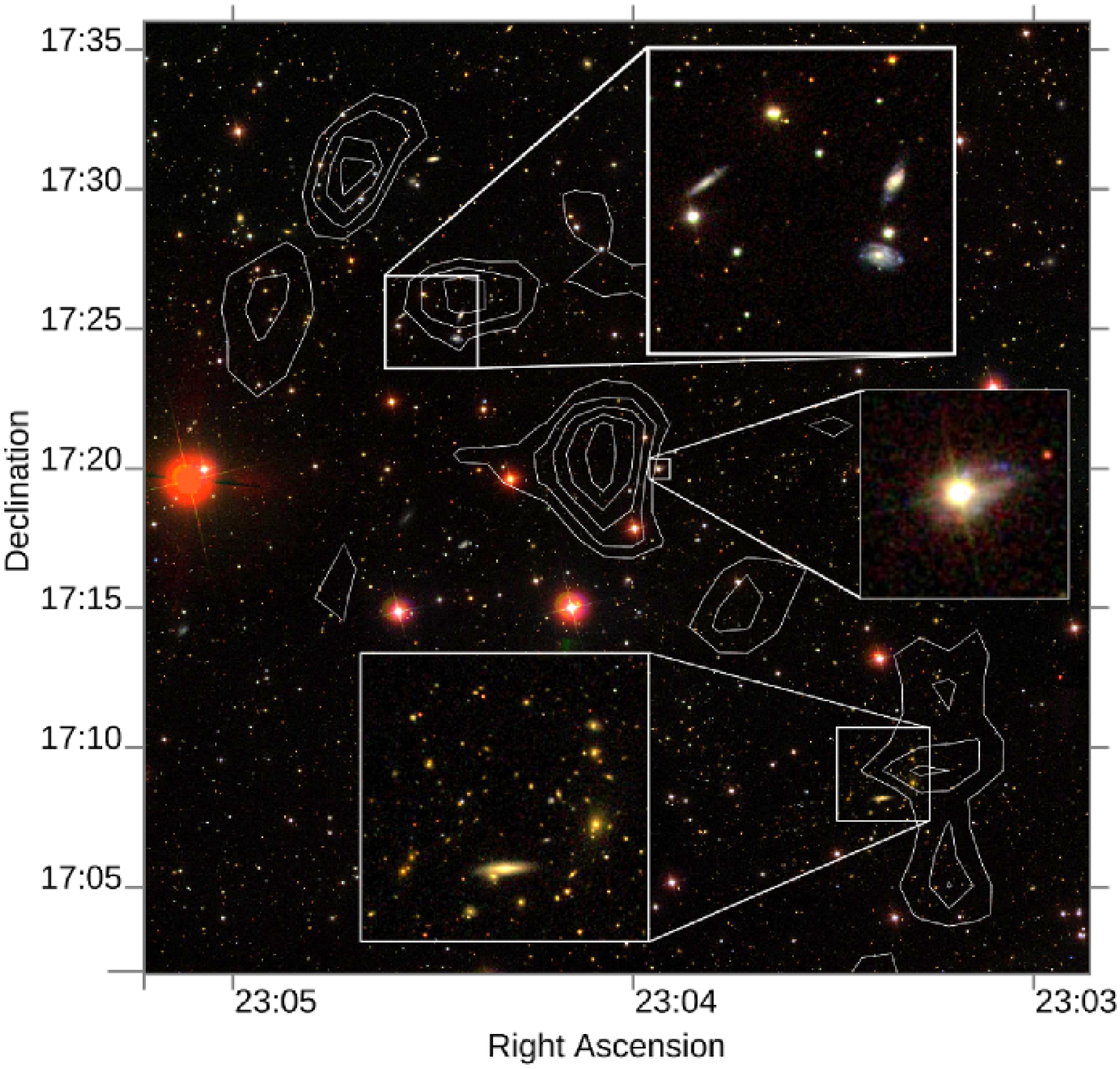}} 
\caption[252Maps]{Integrated flux contours (over the velocity range 11,025 to 11,614 km~s$^{-1}$) of AF7448\_202 (north), a source with at least three possible optical counterparts including a ring galaxy. Optical counterparts (SDSS RGB images) are shown in the inset images. Contours are from 0.1 to 1.0 in steps of 0.1 Jy/beam km~s$^{-1}$. The green circle indicates the 3.5$^\prime$ Arecibo beam. AF7448\_204 (centre) is optically disturbed and there is some suggestion in the residual map (right panel, see text for details) that its H\textsc{i} is also disturbed. A possible north-south extension is also visible in AF7448\_193 (south).}
\label{fig:ring}
\end{center}
\end{figure*}

Four of the nine galaxies we identify as peculiar or mergers show evidence of extended H\textsc{i} emission, but of those, in three cases the H\textsc{i} components overlap those of other galaxies. Only one shows unequivocal evidence of H\textsc{i} extended into intergalactic space -- AF7448\_085, which we identify as UGC12342. This is a spectacular merger but does not (from a NED search) appear to have any previous H\textsc{i} mapping. We will discuss this in detail in a future paper, but for now we only note that we detect H\textsc{i} emission extending over 13$^\prime$ (about 450 kpc in projected extent).

The objects which we are unable to classify are most likely late-type galaxies, as evidenced by their blue colours, but their small angular size (typically $<$ 0.3$^\prime$) makes it difficult to identify any spiral, irregular or other structures.

The early-type galaxies are, as expected, amongst the reddest objects in the sample. There is no clear correlation between their M$_{HI}$/L$_{g}$ ratio and luminosity, however, all are relatively gas rich compared to late-type objects of similar colour. In contrast, the optically bright early-types we detected in H\textsc{i} in the Virgo cluster (paper V) were all gas poor. This suggests that the origin of the gas in early-types may vary between the cluster and field population, and is also consistent with the idea that some of the Virgo S0s may be spiral galaxies which have lost nearly all of their original gas content.

One notable example is AF7448\_073, which we identify as CGCG 453-069. With M$_{g}$ = -20.4 and $g$-$i$ = 1.4, this is one of the brightest and reddest objects in our sample. Unfortunately at this luminosity the blue and red sequences are indistinguishable, but morphologically there is no doubt that this is an early-type object. Surprisingly, its M$_{HI}$/L$_{g}$ ratio of 0.2 is not particularly low given its luminosity. It seems unlikely that objects like this could acquire such large amounts of gas (the total H\textsc{i} content is 6.1$\times$10$^{9}$ M$_{\odot}$) without triggering star formation. It is equally unlikely that the gas could remain stable against star formation in an environment such as this.

\subsection{Non-detections}
A NED search finds 58 galaxies with optical redshift data available in this region (with $cz$ $<$ 20,000 km~s$^{-1}$) and no previous H\textsc{i} measurements. However 19 of these are in such close proximity to H\textsc{i} detections that we cannot rule out their (at least partial) contribution to some of our sources. This leaves a sample of 39. We examine the H\textsc{i} spectra for each of these from our data cube, confirming that these are indeed non-detections and not just missed by our source extraction procedures. We classify 24 of these as early-type galaxies, 8 as spirals and 7 unknown.

The non-detection of these galaxies is unsurprising. Their varying distances give them H\textsc{i} sensitivity limits ranging from 1$\times$10$^{9}$ M$_{\odot}$ to 1$\times$10$^{10}$ M$_{\odot}$. M$_{HI}$/L$_{g}$ ratio sensitivity limits range from 0.3 to 2.2, which would be entirely consistent with other objects of their luminosity. An H\textsc{i} mass even slightly lower than normal (which would still be well within the observed scatter) would render them undetectable.

\section{Summary and discussion}
\label{sec:Conclusions}
We have described an environment dominated by extragalactic filaments with a high galaxy density. As many as 79 galaxies (in 34 separate structures) show signs of extended H\textsc{i} emission. Approximately half of these may be merely the result of source confusion due to the high galaxy density in this region. Another third are more ambiguous cases where the extension may result from faint companion galaxies (though lack of optical redshift data prevents us from determining if these are really associated with the extension). The remaining cases extend at least one beam width into intergalactic space with no plausible optical counterparts visible in the SDSS.

Several of the galaxies with possible extensions are obvious ongoing mergers, but most are relatively normal galaxies. Unlike galaxies in clusters, the galaxies here do not have measurably different colours or gas contents compared to those in less dense environments. The prevalence of extended features does not appear to relate to any abnormal trends within the sample. The galaxies here have, however, also lost much less gas than in environments like Virgo, where galaxies are clearly redder than in the field (see paper V, for example).

We find that galaxies generally have low relative velocities ($<$ 200 km~s$^{-1}$) in systems where extended H\textsc{i} is detected. This is consistent with the velocity dispersion of the large-scale structures, after subtracting their overall velocity gradients. The amount of extended gas is in some cases comparable to (but rarely exceeds) the amount detected within its presumed progenitor galaxies and is often only a small fraction ($<$ 20\%) of the total gas in the system.
 
Of the nine mergers we identified, six are very blue given their luminosity and gas content, perhaps suggesting merger-triggered star formation. In contrast, gas-rich ($>$10$^{9}$ M$_{\odot}$) elliptical objects with disturbed H\textsc{i} components suggest that neither accretion of gas nor tidal interactions necessarily \textit{always} trigger star formation - the gas can remain stable against it. Yet we detect not a single case of H\textsc{i} where it is not clearly associated with an optical component. The H\textsc{i} streams invariably terminate in an optical galaxy (at least at one end and sometimes both).

Accretion of gas from the intergalactic medium could explain how the local star formation rate has remained constant over the lifetime of the Milky Way (\citealt{bin2000}) despite the fact that at the present rate the gas should be exhausted within much less than a Hubble time (\citealt{ken}). Accretion may be a more logical explanation for the chains of galaxies with possible links by multiple H\textsc{i} streams that we detect here (see \ref{fig:MomMap1}-\ref{fig:MomMap3}) than tidal interactions - in this scenario, the filamentary substructures we see would result from the gas tracing the underlying CDM structure.

Distinguishing between accretion and removal of gas is not straightforward. The quantities of intergalactic gas are in many cases relatively low compared to the intragalactic H\textsc{i} - so we expect the effects of gas removal to be subtle. There are, however, several galaxies in which the case for a recent merger is unequivocal (UGC12342 being the prime example) which show clear differences in their optical properties. Furthermore, the high density and low relative velocities of many galaxies strongly suggests that tidal interactions \textit{must} be occurring, regardless of any other process which may be at work.

One signature feature resulting from accretion is the presence of extra-planar gas, since it is much easier if the interaction occurs in the plane of rotation. Some of our detections do show evidence of extended emission outside the plane of the disc (see figure \ref{fig:ring}, for example). This suggests several different processes are at work in this environment, possibly with accretion and gas removal occurring simultaneously. The large variation in the relative proportions of gas in the streams compared to the galaxies also exemplifies that understanding the processes occurring is not straightforward.

It appears we are dealing with an environment where there are many individually unusual systems, but few common trends between them. There are clearly many tidal interactions occurring, but few mergers. There is a significant amount of intergalactic H\textsc{i}, which has often been regarded as the eventual source of the hot X-ray gas found in galaxy clusters. The low velocity dispersion of the large-scale structures seen here make it unlikely that we are witnessing the formation of a cluster. Certainly, however, we see that there is significant gas loss and even morphological evolution within some of the galaxy systems detected, and this pre-processing is often regarded as an important part of cluster evolution (e.g. \citealt{bg06}). The galaxies in this study may not be demonstrating preprocessing directly - there is no suggestion they will ever form a cluster - but they do show that evolution in the field can be important.

\section*{Acknowledgements}

This work is based on observations collected at Arecibo Observatory. The Arecibo Observatory is operated by SRI International under a cooperative agreement with the National Science Foundation (AST-1100968), and in alliance with Ana G. M\'endez---Universidad Metropolitana, and the Universities Space Research Association. 

This work was supported by the project RVO:67985815.

This research has made use of the NASA/IPAC Extragalactic Database (NED) which is operated by the Jet Propulsion Laboratory, California Institute of Technology, under contract with the National Aeronautics and Space Administration. 

This work has made use of the SDSS. Funding for the SDSS and SDSS-II has been provided by the Alfred P. Sloan Foundation, the Participating Institutions, the National Science Foundation, the U.S. Department of Energy, the National Aeronautics and Space Administration, the Japanese Monbukagakusho, the Max Planck Society, and the Higher Education Funding Council for England. The SDSS Web Site is http://www.sdss.org/.

The SDSS is managed by the Astrophysical Research Consortium for the Participating Institutions. The Participating Institutions are the American Museum of Natural History, Astrophysical Institute Potsdam, University of Basel, University of Cambridge, Case Western Reserve University, University of Chicago, Drexel University, Fermilab, the Institute for Advanced Study, the Japan Participation Group, Johns Hopkins University, the Joint Institute for Nuclear Astrophysics, the Kavli Institute for Particle Astrophysics and Cosmology, the Korean Scientist Group, the Chinese Academy of Sciences (LAMOST), Los Alamos National Laboratory, the Max-Planck-Institute for Astronomy (MPIA), the Max-Planck-Institute for Astrophysics (MPA), New Mexico State University, Ohio State University, University of Pittsburgh, University of Portsmouth, Princeton University, the United States Naval Observatory, and the University of Washington.

{}

\appendix
\section{FRELLED}
\label{sec:apsource}

There are two key features to \textsc{frelled} which allow for much more rapid visual source extraction than was previously possible. The first is the ability for the user to freely rotate the view of a data cube, exploring its true 3-dimensional structure, in realtime. Other viewers, such as \textsc{ds9} (v. 7.2), require a substantial delay (tens of seconds) between rotation and the view updating, with ours the view is updated instantaneously as the viewpoint is altered (the older \textit{xray} (\citealt{xray}) program is much faster than \textsc{ds9}, but has a more complex interface and no facilities to aid in source cataloguing), hence its name.

The second feature is the ability of the user to mask sources interactively. These 3D masks can be named by the user for easy re-location and analysis. They also allow the user to generate input parameters for \textit{mbspect} (see section \ref{sec:Analysis}), create moment maps, and query NED and the SDSS. \textsc{frelled} is written as a series of Python scripts for the graphics suite, ``Blender'', and the source code is available on our website (www.naic.edu/$\sim$ages). For a thorough review of Blender for astronomers, see \citealt{bkent}.

\begin{figure}
\begin{center}  
  \subfloat[]{\includegraphics[height=64mm]{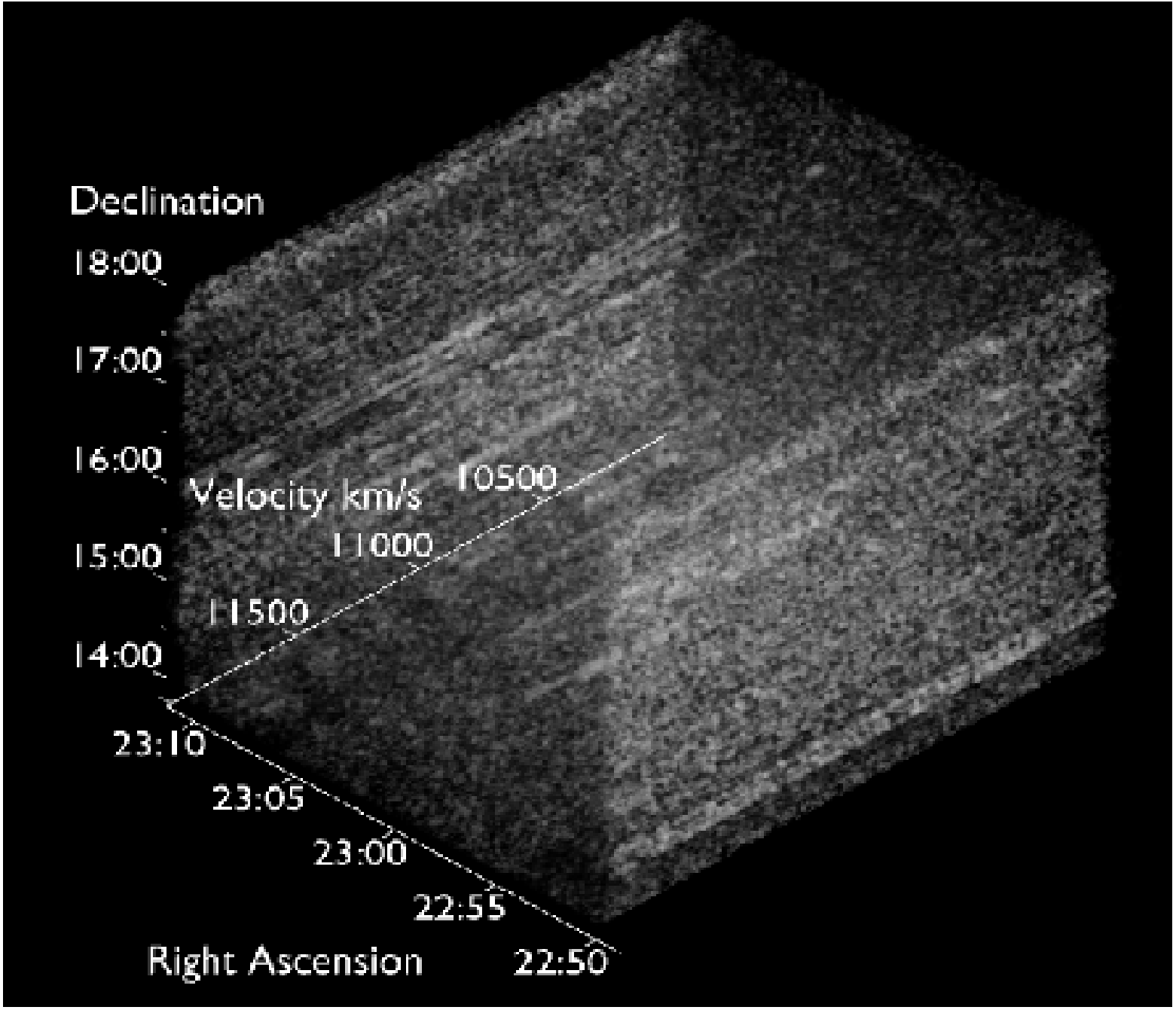}}\\
  \subfloat[]{\includegraphics[height=64mm]{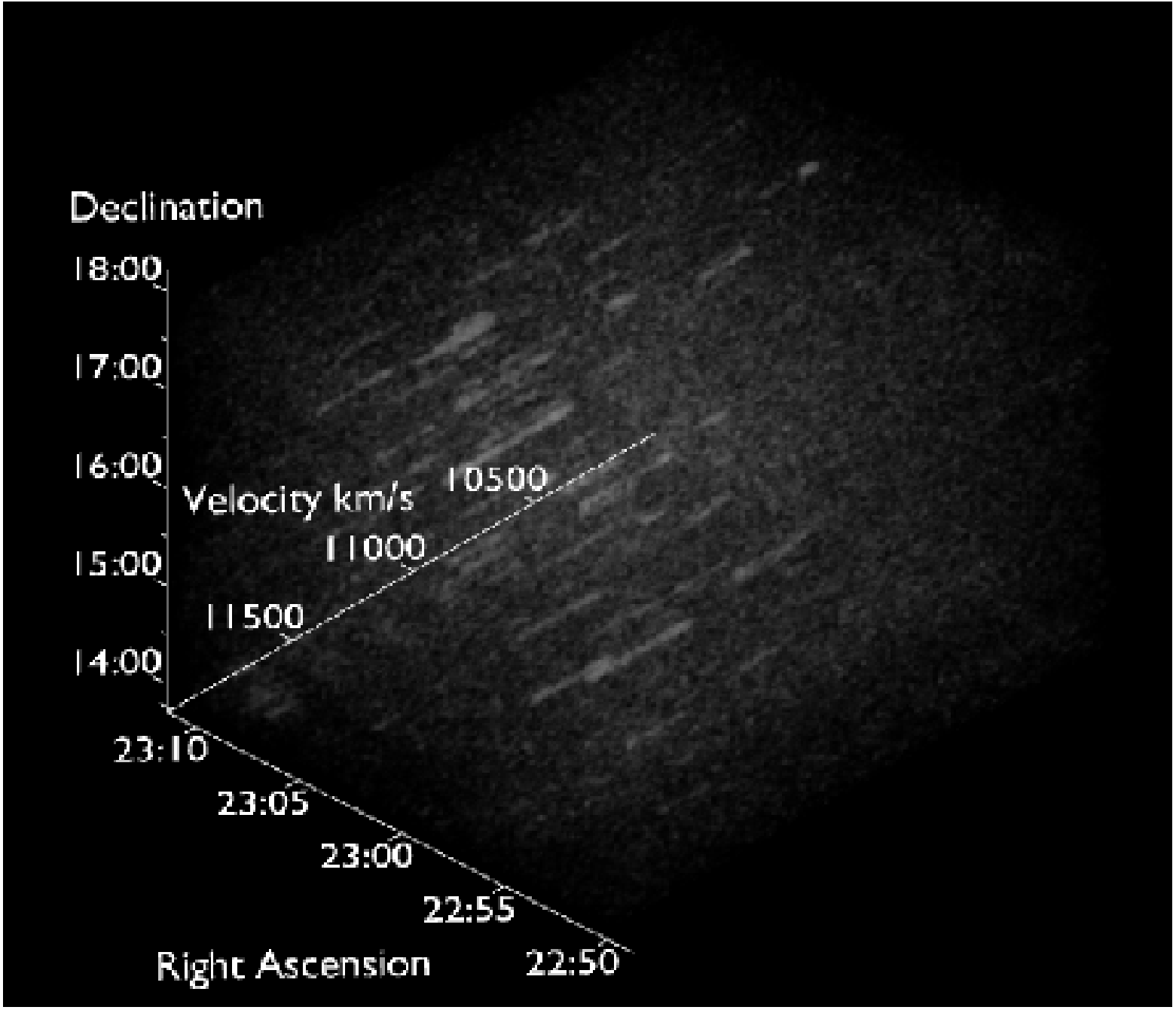}} 
\caption[FluxSN3D]{Views of the NGC 7448 data cube from \textsc{frelled}, showing the effects of (a) using flux values to determine the colours (b) using S/N values to determine the colours. We note that despite the dramatic differences between the two images, both display approximately equivalent intensity ranges. One could (potentially) detect sources of the same flux by using either display method - it is simply that much of the noise visible in (a) has been removed in (b), making source extraction far easier. A rotating movie of this figure in included in the online material.}
\label{fig:FluxSN}
\end{center}
\end{figure}

Viewing AGES data in 3D can be problematic since the noise level is much higher at the spatial edges of the cube, where sensitivity is lower. This means that there is a much greater range of flux values for pixels in these regions compared to those in the fully-sampled area. Using a fixed intensity range based on flux causes the much brighter pixels at the edges to block the view of the interior of the cube. To avoid removing these regions altogether, we recast the cube from flux to S/N. We divide the  flux value of each pixel by the sigma-clipped \textit{rms} estimated along the spectrum containing it. Since the noise is still Gaussian in the regions of higher \textit{rms}, the S/N range is the same as in any other region of the cube. The results are shown in figure \ref{fig:FluxSN}.

Converting the cube into S/N obviously renders the pixel values unphysical. It is also more difficult to display faint features in this way. The pixel colour is calculated by summing the values along the line of sight (equivalent to moment 0), so faint features tend to be overwhelmed when there is significant variation in the noise. Calculating the colour based on the maximum value along the line of sight (moment -2) would be better, but would still not help for the faintest detectable features. Instead we also use a 2D variant of \textsc{frelled}, in which we can view individual slices of the cube in a style similar to \textit{kvis} (\citealt{kvis}). Our source extraction is thus a two-stage process : first, we search a 3D version of the cube where we can very quickly mask the bright sources, then, we import these masks into the 2D viewer and search more carefully for fainter sources.

\section{H\textsc{i} maps}
\begin{figure*}
\includegraphics[width=170mm]{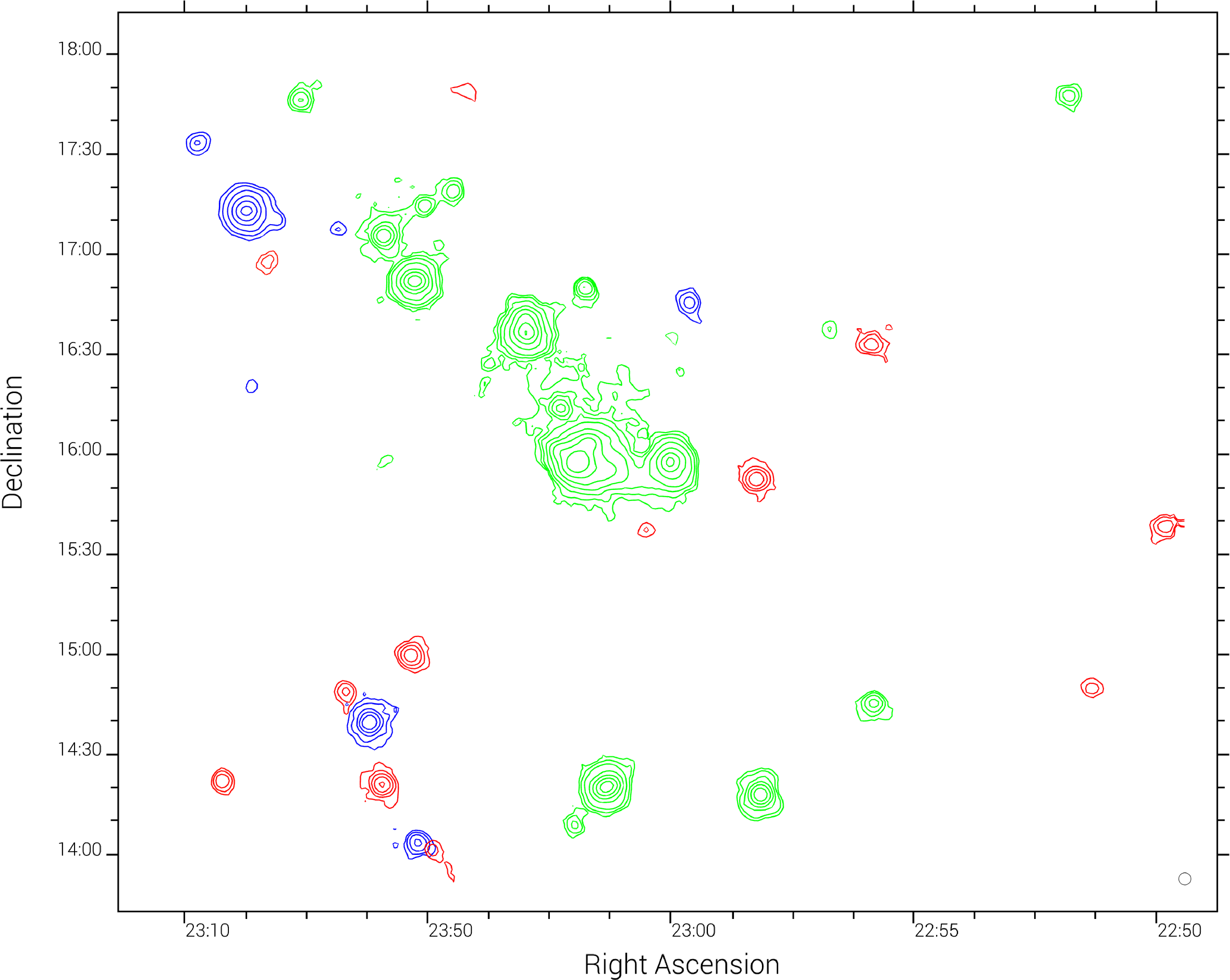}
\nopagebreak
\caption[100 - 5,000 km~s$^{-1}$ Moment map]{Moment maps for sources detected in velocity range 100 $<$ $cz$ $5,000$ km~s$^{-1}$. The minimum contour level is 0.1 Jy~beam$^{-1}$ km~s$^{-1}$ with further levels each a factor of two higher than the previous. Colours indicate velocity - blue contours are detections from 100-1,600 km~s$^{-1}$, green are from 1,600-3,200 km~s$^{-1}$, and red are from 3,200-5,000 km~s$^{-1}$. The Arecibo beam is shown as a black circle in the lower-right.}
\label{fig:MomMap1}
\end{figure*}

\begin{figure*}
\includegraphics[width=170mm]{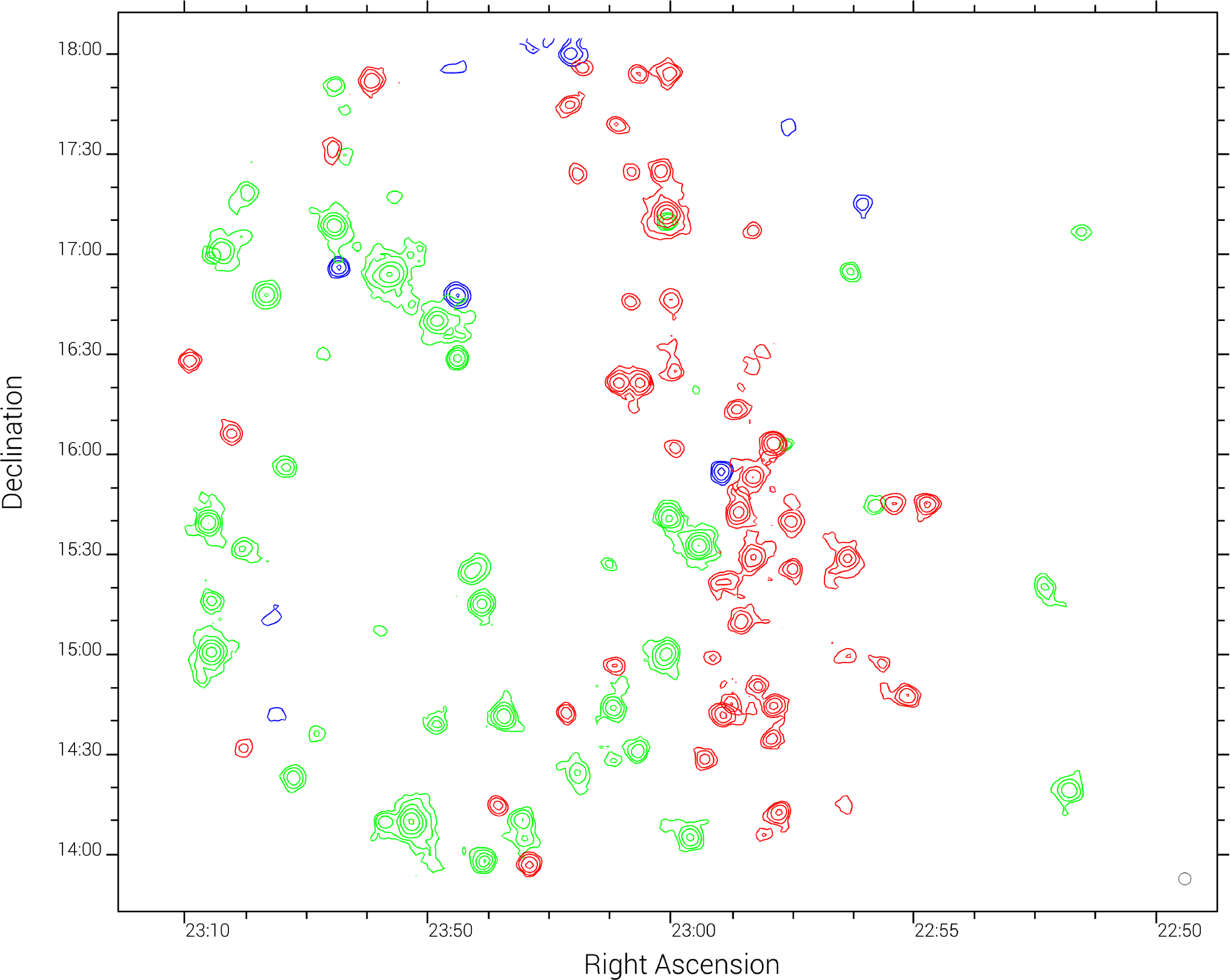}
\caption[5000 - 10,000 km~s$^{-1}$ Moment map]{Moment maps for sources detected in velocity range 5,000 $<$ $cz$ $10,000$ km~s$^{-1}$. The minimum contour level is 0.1 Jy~beam$^{-1}$ km~s$^{-1}$ with further levels each a factor of two higher than the previous. Colours indicate velocity - blue contours are detections from 5,000-6,600 km~s$^{-1}$, green are from 6,600-8,700 km~s$^{-1}$, and red are from 8,700-10,000 km~s$^{-1}$. The Arecibo beam is shown as a black circle in the lower-right.}
\label{fig:MomMap2}
\end{figure*}

\begin{figure*}
\includegraphics[width=170mm]{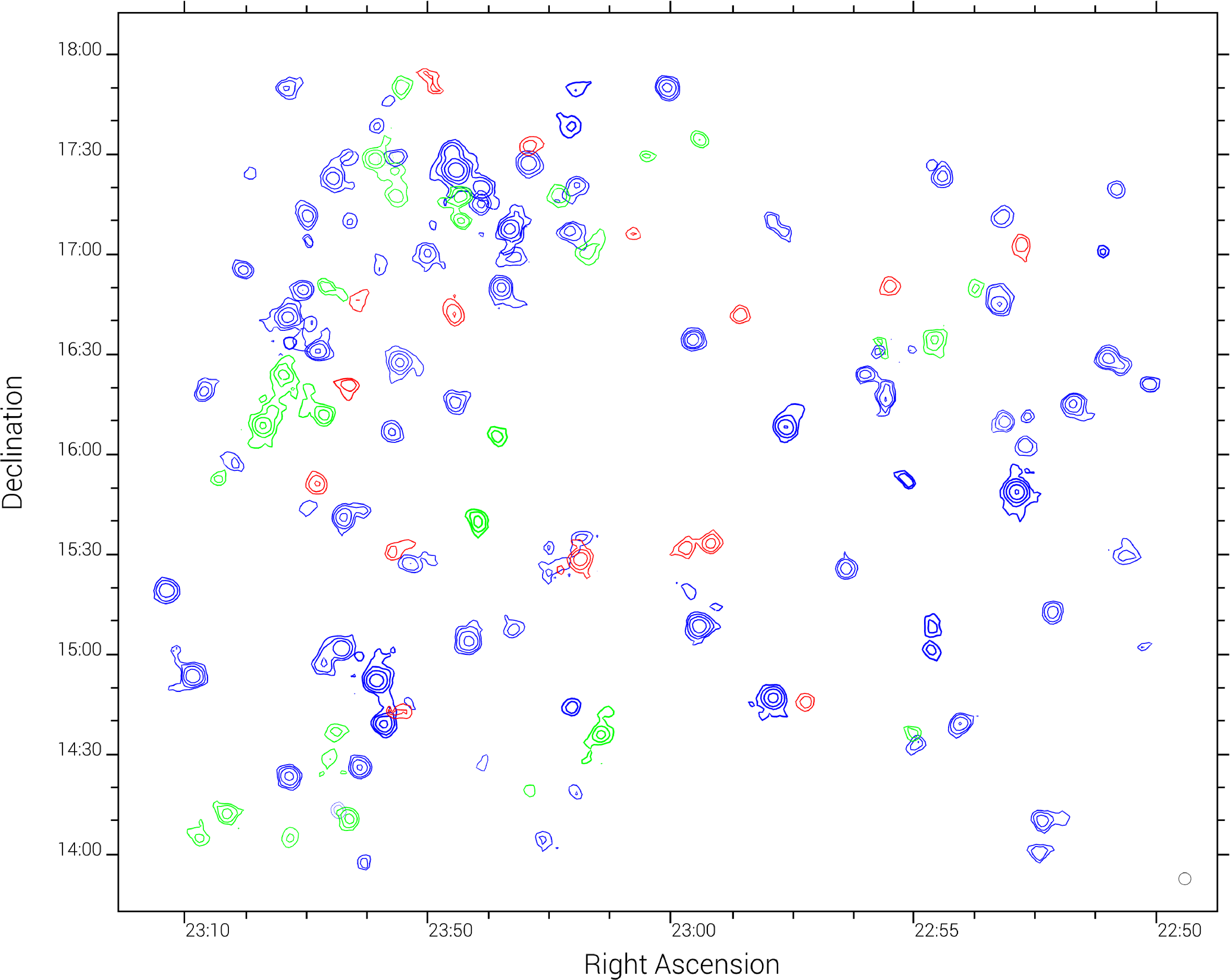}
\caption[10000 - 15,000 km~s$^{-1}$ Moment map]{Moment maps for sources detected in velocity range 10,000 $<$ $cz$ $15,000$ km~s$^{-1}$. The minimum contour level is 0.1 Jy~beam$^{-1}$ km~s$^{-1}$ with further levels each a factor of two higher than the previous. Colours indicate velocity - blue contours are detections from 10,000-11,600 km~s$^{-1}$, green are from 11,600-13,700 km~s$^{-1}$, and red are from 13,700-15,000 km~s$^{-1}$. The Arecibo beam is shown as a black circle in the lower-right.}
\label{fig:MomMap3}
\end{figure*}

\label{lastpage}

\end{document}